\documentclass[11pt]{article}
\usepackage[mathscr]{eucal}
\usepackage{mathrsfs}
\usepackage{amssymb}
\usepackage{calligra}
\usepackage{fontenc}
\usepackage{amsbsy} 
\usepackage{graphicx}
\graphicspath{%
    {converted_graphics/}
    {/}
}
\begin{document}

\newcommand{\ts}{{\tilde{\sf s}}}
\newcommand{\sfv}{{\sf v}}
\newcommand{\sfw}{{\sf w}}
\newcommand{\simge}{\ba{cc}\vspace*{-2.4mm}>\\ \sim\ea }
\newcommand{\simle}{\ba{cc}\vspace*{-2.4mm}<\\ \sim\ea }
\newcommand{\Cdot}{\!\cdot\!}
\newcommand{\sq}{{$\sqcap\!\!\!\!\sqcup$}}
\newcommand{\Eu}{{\rm I\,\!\! E}}
\newcommand{\Io}{\Int{\Omega}{}}
\newcommand{\Id}{\Int{\cald}{}}
\newcommand{\Div}{\mbox{\rm div}\,}
\newcommand{\tr}{\mbox{\rm tr}\,}
\newcommand{\grad}{\mbox{\rm grad}\,}
\newcommand{\supp}{\mbox{\rm supp}\,}
\newcommand{\curl}{\mbox{\rm curl}\,}
\newcommand{\Ido}{\Int{\partial\Omega}{}}
\newcommand{\IdS}{\Int{\Sigma}{}}
\newcommand{\Oint}[2]{{\displaystyle \oint_{#1}^{#2}}}
\newcommand{\Int}[2]{{\displaystyle \int_{ #1}^{ #2}}}
\newcommand{\Lim}[1]{{\displaystyle \lim_{ #1}}}
\newcommand{\Limsup}[1]{{\displaystyle \limsup_{\footnotesize #1}}}
\newcommand{\Liminf}[1]{{\displaystyle \liminf_{\footnotesize #1}}}
\newcommand{\Sup}[1]{{\displaystyle \sup_{#1}}}
\newcommand{\Inf}[1]{{\displaystyle \inf_{#1}}}
\newcommand{\Max}[1]{{\displaystyle \max_{#1}}}
\newcommand{\Min}[1]{{\displaystyle \min_{#1}}}
\newcommand{\Sum}[2]{{\displaystyle \sum_{#1}^{#2}}}
\newcommand{\Prod}[2]{{\displaystyle \prod_{#1}^{#2}}}
\newcommand{\BCup}[2]{{\displaystyle \bigcup_{#1}^{#2}}}
\newcommand{\BCap}[2]{{\displaystyle \bigcap_{#1}^{#2}}}
\newcommand{\Frac}[2]{\displaystyle{\frac{\displaystyle{#1}}{\displaystyle{#2}}}}
\newcommand{\norm}[1]{\left\|{#1}\right\|}
\newcommand{\Norm}[1]{\langle\langle{#1}\rangle\rangle_q}
\newcommand{\No}[1]{\langle\!\langle{#1}\rangle\!\rangle}
\newcommand{\NO}[1]{{\langle{#1}\rangle}_{\lambda,q}}
\newcommand{\beea}{\begin{eqnarray}}
\newcommand{\eeea}{\end{eqnarray}}
\newcommand{\ms}{\medskip\smallskip}
\newcommand{\bs}{\bigskip}
\newcommand{\ps}{\par\smallskip}
\newcommand{\bfe}{{\mbox{\boldmath $e$}} }
\newcommand{\pni}{{\par\noindent}}
\newcommand{\bfq}{{\mbox{\boldmath $q$}} }
\newcommand{\bfz}{{\mbox{\boldmath $z$}} }
\newcommand{\0}{{\mbox{\boldmath $0$}} }
\newcommand{\LE}{\!\!\!&\le&\!\!\!}
\newcommand{\BL}[1]{{\par\smallskip{\bf Lemma #1.}}}
\newcommand{\BT}[1]{{\par\smallskip{\bf Theorem #1.}}}
\newcommand{\Ln}{[\!|}
\newcommand{\Rn}{|\!]}
\newcommand{\n}[1]{{\Ln{#1}\Rn}} 
\newcommand{\nq}[1]{{\Ln{#1}\Rn}_{q}} 
\newcommand{\nqr}[1]{{\Ln{#1}\Rn}_{q,r}} 
\newcommand{\Nq}[1]{{\langle{#1}\rangle}_{q}} 
\newcommand{\Nql}[1]{{\langle{#1}\rangle}_{\lambda,q}} 
\newcommand{\Nqr}[1]{{\langle{#1}\rangle}_{q,r}}
\newcommand{\N}[1]{{|\!\!|\!\!|\,{#1}\,|\!|\!\!|_2}}
\newcommand{\EA}[2]{$$#1$$%
\vspace{-6.mm}
\begin{equation}
\end{equation}
\vspace{-6.mm}
$$
#2
\setlength{\belowdisplayskip}{3mm}
\setlength{\belowdisplayshortskip}{3mm}
$$
}
\newcommand{\A}[2]{$$#1$$%
\vspace{-4.mm}
$$
#2
\setlength{\belowdisplayskip}{3mm}
\setlength{\belowdisplayshortskip}{3mm}
$$
}
\newcommand{\BF}{\begin{footnotesize}}
\newcommand{\EF}{\end{footnotesize}}
\setlength{\jot}{.15in}
\newcommand{\pde}[2]{{\displaystyle \frac{\mbox{$\partial #1$}}{\mbox{$\partial #2$}}}}
\newcommand{\ode}[2]{{\displaystyle \frac{\mbox{$d #1$}}{\mbox{$d #2$}}}}
\newcommand{\f}[2]{\frac{\mbox{$#1$}}{\mbox{$ #2$}}}
\newcommand{\bi}{\begin{itemize}}
\newcommand{\ei}{\end{itemize}}
\newcommand{\ed}{\end{document}}
\newcommand{\be}{\begin{equation}}
\newcommand{\ba}{\begin{array}}
\newcommand{\ea}{\end{array}}
\newcommand{\ee}{\end{equation}}
\newcommand{\eeq}[1]{\label{eq:#1}\end{equation}}
\newcommand{\real}{{\mathbb R}}
\newcommand{\compl}{{\mathbb C}}
\def\Id{\mbox{\boldmath $1$}}
\def\zero{\mbox{\boldmath $0$}}
\newcommand{\PP}{{\rm I\!\!\,P}}
\newcommand{\nat}{{\mathbb N}}
\newcommand{\bfpsi}{\mbox{\boldmath $\psi$}}
\newcommand{\bfchi}{\mbox{\boldmath $\chi$}}
\newcommand{\bfomega}{\mbox{\boldmath $\omega$}}
\newcommand{\bfvaromega}{\mbox{\boldmath $\varpi$}}
\newcommand{\bfOmega}{\mbox{\boldmath $\Omega$}}
\newcommand{\bfTheta}{\mbox{\boldmath $\Theta$}}
\newcommand{\bfxi}{\mbox{\boldmath $\xi$}}
\newcommand{\bfmu}{\mbox{\boldmath $\mu$}}
\newcommand{\bfx}{\mbox{\boldmath $x$}}
\newcommand{\bfy}{\mbox{\boldmath $y$}}
\newcommand{\bfPsi}{\mbox{\boldmath $\Psi$}}
\newcommand{\bfphi}{\mbox{\boldmath $\varphi$}}
\newcommand{\bfhi}{\mbox{\boldmath $\phi$}}
\newcommand{\bfPhi}{\mbox{\boldmath $\Phi$}}
\newcommand{\bfv}{{\mbox{\boldmath $v$}} }
\newcommand{\bfu}{{\mbox{\boldmath $u$}} }
\newcommand{\bfsf}{{\mbox{\footnotesize\boldmath $s$}} }
\newcommand{\bfuf}{{\mbox{\footnotesize\boldmath $u$}} }
\newcommand{\bfw}{{\mbox{\boldmath $w$}} }
\newcommand{\bff}{{\mbox{\boldmath $f$}} }
\newcommand{\bfa}{{\mbox{\boldmath $a$}} }
\newcommand{\bfi}{{\mbox{\boldmath $i$}} }
\newcommand{\bfj}{{\mbox{\boldmath $j$}} }
\newcommand{\bfc}{{\mbox{\boldmath $c$}} }
\newcommand{\bfo}{{\mbox{\boldmath $o$}} }
\newcommand{\bfp}{{\mbox{\boldmath $p$}} }
\newcommand{\bfkp}{{\mbox{\footnotesize{\boldmath $k$}}} }
\newcommand{\bfka}{{\mbox{\footnotesize{\boldmath $k^*$}}} }
\newcommand{\bft}{{\mbox{\boldmath $t$}} }
\newcommand{\bfd}{{\mbox{\boldmath $d$}} }
\newcommand{\bfl}{{\mbox{\boldmath $l$}} }
\newcommand{\bfr}{{\mbox{\boldmath $r$}} }
\newcommand{\bfk}{{\mbox{\boldmath $k$}} }
\newcommand{\bfA}{{\mbox{\boldmath $A$}} }
\newcommand{\bfS}{{\mbox{\boldmath $S$}} }
\newcommand{\bfO}{{\mbox{\boldmath $O$}} }
\newcommand{\bfM}{{\mbox{\boldmath $M$}} }
\newcommand{\bfP}{{\mbox{\boldmath $P$}} }
\newcommand{\bfB}{{\mbox{\boldmath $B$}} }
\newcommand{\bfR}{{\mbox{\boldmath $R$}} }
\newcommand{\bfC}{{\mbox{\boldmath $C$}} }
\newcommand{\bfD}{{\mbox{\boldmath $D$}} }
\newcommand{\bfQ}{{\mbox{\boldmath $Q$}} }
\newcommand{\bfZ}{{\mbox{\boldmath $Z$}} }
\newcommand{\bfG}{{\mbox{\boldmath $G$}} }
\newcommand{\bfE}{{\mbox{\boldmath $E$}} }
\newcommand{\bfX}{{\mbox{\boldmath $X$}} }
\newcommand{\bfY}{{\mbox{\boldmath $Y$}} }
\newcommand{\bfH}{{\mbox{\boldmath $H$}} }
\newcommand{\bfI}{{\mbox{\boldmath $I$}} }
\newcommand{\bfJ}{{\mbox{\boldmath $J$}} }
\newcommand{\bfN}{{\mbox{\boldmath $N$}} }
\newcommand{\bfh}{{\mbox{\boldmath $h$}} }
\newcommand{\bfm}{{\mbox{\boldmath $m$}} }
\newcommand{\bfone}{{\mbox{\boldmath $1$}} }
\newcommand{\hs}{{\rm I}\!\!\,{\rm R}^3_+}
\newcommand{\cala}{{\cal A}}
\newcommand{\calb}{{\cal B}}
\newcommand{\calc}{{\cal C}}
\newcommand{\cald}{{\cal D}}
\newcommand{\cale}{{\cal E}}
\newcommand{\calf}{{\cal F}}
\newcommand{\calg}{{\cal G}}
\newcommand{\calh}{{\cal H}}
\newcommand{\cali}{{\cal I}}
\newcommand{\calj}{{\cal J}}
\newcommand{\calk}{{\cal K}}
\newcommand{\call}{{\cal L}}
\newcommand{\calm}{{\cal M}}
\newcommand{\caln}{{\cal N}}
\newcommand{\calo}{{\cal O}}
\newcommand{\calp}{{\cal P}}
\newcommand{\calq}{{\cal Q}}
\newcommand{\calr}{{\cal R}}
\newcommand{\cals}{{\cal S}}
\newcommand{\calt}{{\cal T}}
\newcommand{\calu}{{\cal U}}
\newcommand{\calv}{{\cal V}}
\newcommand{\calx}{{\cal X}}
\newcommand{\caly}{{\cal Y}}
\newcommand{\calw}{{\cal W}}
\newcommand{\calz}{{\cal Z}}
\newcommand{\bfsigma}{\mbox{\boldmath $\sigma$}}
\newcommand{\bfSigma}{\mbox{\boldmath $\Sigma$}}
\newcommand{\bftau}{\mbox{\boldmath $\tau$}}
\newcommand{\bfeta}{\mbox{\boldmath $\eta$}}
\newcommand{\bfT}{{\mbox{\boldmath $T$}} }
\newcommand{\bfV}{{\mbox{\boldmath $V$}} }
\newcommand{\bfU}{{\mbox{\boldmath $U$}} }
\newcommand{\bfW}{{\mbox{\boldmath $W$}} }
\newcommand{\bfF}{{\mbox{\boldmath $F$}} }
\newcommand{\bfK}{{\mbox{\boldmath $K$}} }
\newcommand{\bfL}{{\mbox{\boldmath $L$}} }
\newcommand{\bfb}{{\mbox{\boldmath $b$}} }
\newcommand{\bfg}{{\mbox{\boldmath $g$}} }
\newcommand{\bfn}{{\mbox{\boldmath $n$}} }
\newcommand{\bfs}{{\mbox{\boldmath $s$}} }
\newcommand{\cf}{{\it cf.} }
\newcommand{\io}{\int_\Omega}
\newcommand{\1}{\item[({\it i})]}
\newcommand{\2}{\item[({\it ii})]}
\newcommand{\3}{\item[({\it iii})]}
\newcommand{\4}{\item[({\it iv})]}
\newcommand{\5}{\item[({\it v})]}
\newcommand{\6}{\item[({\it vi})]}
\newcommand{\7}{\item[({\it vii})]}
\newcommand{\8}{\item[({\it viii})]}
\newcommand{\9}{\item[({\it xi})]}
\newcommand{\ido}{\int_{\partial\Omega}}
\newcommand{\half}{\mbox{$\frac{1}{2}$}}
\def\parallel{\|}
\def\mid{|}
\def\Bbb R{\real}
\def\hat{\widehat}
\def\tilde{\widetilde}
\def\bar{\overline}
\newcommand{\threehalves}{3\over 2}
\newcommand{\bfPi}{\mbox{\boldmath $\Pi$}}
\newcommand{\bfXi}{\mbox{\boldmath $\Xi$}}
\newcommand{\bfalpha}{\mbox{\boldmath $\alpha$}}
\newcommand{\bfbeta}{\mbox{\boldmath $\beta$}}
\newcommand{\bfgamma}{\mbox{\boldmath $\gamma$}}
\newcommand{\bfdelta}{\mbox{\boldmath $\delta$}}
\newcommand{\bfzeta}{\mbox{\boldmath $\zeta$}}
\newcommand{\bfUpsilon}{\mbox{\boldmath $\Upsilon$}}
\newcommand{\bfGamma}{\mbox{\boldmath $\Gamma$}}
\newcommand{\bfcala}{\mbox{\boldmath ${\cal A}$}}
\newcommand{\bfcalm}{\mbox{\boldmath ${\cal M}$}}
\newcommand{\bfcaln}{\mbox{\boldmath ${\cal N}$}}
\newcommand{\bfcalq}{\mbox{\boldmath ${\cal Q}$}}
\newcommand{\bfcalb}{\mbox{\boldmath ${\cal B}$}}
\newcommand{\bfcalc}{\mbox{\boldmath ${\cal C}$}}
\newcommand{\bfcali}{\mbox{\boldmath ${\cal I}$}}
\newcommand{\bfcalg}{\mbox{\boldmath ${\cal G}$}}
\newcommand{\bfcalh}{\mbox{\boldmath ${\cal H}$}}
\newcommand{\bfcalk}{\mbox{\boldmath ${\cal K}$}}
\newcommand{\bfcalt}{\mbox{\boldmath ${\cal T}$}}
\newcommand{\bfcalx}{\mbox{\boldmath ${\cal X}$}}
\newcommand{\bfcall}{\mbox{\boldmath ${\cal L}$}}
\newcommand{\bfcalf}{\mbox{\boldmath ${\cal F}$}}
\newcommand{\bfcalr}{\mbox{\boldmath ${\cal R}$}}
\newcommand{\bfcals}{\mbox{\boldmath ${\cal S}$}}
\newcommand{\bfcalw}{\mbox{\boldmath ${\cal W}$}}
\newcommand{\bfcalu}{\mbox{\boldmath ${\cal U}$}}
\newcommand{\bfcalv}{\mbox{\boldmath ${\cal V}$}}
\newcommand{\bfcalz}{\mbox{\boldmath ${\cal Z}$}}
\pagenumbering{roman}
\newcommand{\art}[6]{{\I[{\sc #1,}] {#2}, {\it #3}, {\bf #4}, {#5} {[#6]}}}
\newcommand{\ED}{\end{description}}
\newcommand{\I}{\item }
\newcommand{\ra}{\rm a}
\newcommand{\rb}{\rm b}
\newcommand{\rc}{\rm c}
\newcommand{\Hsp}{{\rm I}\!\!\,{\rm R}^n_+}
\newcommand{\Hsn}{{\rm I}\!\!\,{\rm R}^n_-}
\newcommand{\po}[1]{\mbox{$\displaystyle \frac{\mbox{$\partial #1$}}
{\mbox{$\partial x_{1}$}}$}}
\newcommand{\PO}[1]{\mbox{$\displaystyle \frac{\mbox{$\partial #1$}}
{\mbox{$\partial y_{1}$}}$}}
\newcommand{\OP}{\left(\Delta+2\lambda\PO{}\right)}
\newcommand{\op}{\left(\Delta+2\lambda\po{}\right)}
\newcommand{\ft}[1]{
\Frac{1}{(2\pi)^{n/2}}\Int{{\Bbb R}^{n}}{}e^{i{\bf x}\cdot \bfxi}
#1(\xi)d\xi}
\newcommand{\Ft}[1]{
\Frac{1}{2\pi}\Int{{\Bbb R}^{2}}{}e^{i{x}\cdot \xi}
#1(\xi)d\xi}
\newcommand{\Z}{\item[({\it a})]}
\newcommand{\B}{\item[({\it b})]}
\newcommand{\C}{\item[({\it c})]}
\newcommand{\D}{\item[({\it d})]}
\newcommand{\E}{\item[({\it e})]}
\newcommand{\G}{\item[({\it g})]}
\newcommand{\Š}{\`e}
\newcommand{\…}{\`a}
\newcommand{\•}{\`o}
\newcommand{\—}{\`u}
\newcommand{\}{\`{\i}}
\def\tag{\renewcommand{\theequation}}
\newcommand{\Footnote}{~\footnote}
\newcommand{\ie}{{\it i.e.}}
\newcommand{\dist}{\mbox{\rm dist\,}}
\newcommand{\const}{\mbox{\rm const}}
\newcommand{\trace}{\mbox{\rm trace}}
\newcommand{\Bo}{\par\hfill{$\Box$}\par\noindent}
\newcommand{\Nor}[1]{\langle{#1}\rangle_q}
\newcommand{\vs}{\vspace*{.5cm}\par\noindent}
\newcommand{\Vs}{\vspace*{.6cm}\par\noindent}
\newcommand{\Vvs}{\vspace*{.7cm}\par\noindent}
\newcommand{\VVs}{\vspace*{.8cm}\par\noindent}
\newtheorem{definition}{Definition}[section]
\newcommand{\Bd}{\begin{definition}\begin{rm}}
\newcommand{\Ed}{\end{rm}\end{definition}}
\newtheorem{remark}{Remark}[section]
\newcommand{\Br}{\begin{remark}\begin{rm}}
\newcommand{\Er}{\end{rm}\end{remark}}
\newtheorem{proposition}{Proposition}[section]
\newcommand{\Bp}{\begin{proposition}\begin{sl}}
\newcommand{\EP}[1]{\end{sl}\label{proposition:#1}\end{proposition}}
\newcommand{\propref}[1]{{\rm Proposition \ref{proposition:#1}}}
\newcommand{\Bt}{\begin{theorem}\begin{sl}}
\newcommand{\Et}{\end{sl}\end{theorem}}
\newcommand{\Bl}{\begin{lemma}\begin{sl}}
\newcommand{\El}{\end{sl}\end{lemma}}
\newtheorem{theorem}{Theorem}[section]
\newtheorem{lemma}{Lemma}[section]
\newtheorem{corollary}{Corollary}[section]
\newcommand{\eqref}[1]{{\rm (\ref{eq:#1})}}
\newcommand{\Bc}{\begin{corollary}\begin{sl}}
\newcommand{\Ec}{\end{sl}\end{corollary}}
\newcommand{\ET}[1]{\end{sl}\label{theorem:#1}\end{theorem}}
\newcommand{\EDD}[1]{\end{rm}\label{definition:#1}\end{definition}}
\newcommand{\EL}[1]{\end{sl}\label{lemma:#1}\end{lemma}}
\newcommand{\theoref}[1]{{\rm Theorem \ref{theorem:#1}}}
\newcommand{\ER}[1]{\end{rm}\label{remark:#1}\end{remark}}
\newcommand{\EC}[1]{\end{sl}\label{corollary:#1}\end{corollary}}
\newcommand{\remref}[1]{{\rm Remark \ref{remark:#1}}}
\newcommand{\cororef}[1]{{\rm Corollary \ref{corollary:#1}}}
\newcommand{\lemmref}[1]{{\rm Lemma \ref{lemma:#1}}}
\newcommand{\essup}[1]{{\rm ess}\,{{\displaystyle \sup_{\hspace*{-5mm}{#1}}}}}

\renewcommand{\i}{{\rm i}}

\pagenumbering{arabic}
\newcommand{\QED}{{\par\hfill$\square$\par}}
\renewcommand{\thefootnote}{(\arabic{footnote})}
\title{A Time-Periodic Bifurcation Theorem\\ and its Application to  Navier-Stokes Flow \\ Past an Obstacle} 
\author{ Giovanni P. Galdi 
\thanks{Department of Mechanical Engineering and Materials Science, University of Pittsburgh, PA 15261. 
Work  partially supported by NSF DMS Grant-1311983.}}
\date{}
\maketitle
\begin{abstract} We show an abstract time-periodic bifurcation theorem in Banach spaces. The key point as well as the novelty of the method is to split the original evolution equation into two different coupled equations, one for the time-average of the sought solution and the other for the ``purely periodic'' component. This approach may be particularly useful in studying physical phenomena occurring in unbounded spatial regions. Actually, we furnish  a significant application of the theorem, by providing sufficient conditions for time-periodic bifurcation from a steady-state flow of a Navier-Stokes liquid past a three-dimensional obstacle.  
 \end{abstract}

\renewcommand{\theequation}{\arabic{section}.\arabic{equation}}
\setcounter{section}{0}
\section{Introduction} Time-periodic bifurcation from a steady-state regime is a commonly observed  phenomenon in the dynamics of viscous liquid, for both bounded and unbounded flow; see. e.g. \cite[Section 10.3]{Gui}, \cite[Chapter 3]{Tritton}. As is well-known, it may take place when the magnitude of the driving mechanism, {\sf m} (say), reaches a certain critical value, ${\sf m}_c$. Basically, if ${\sf m}<{\sf m}_c$ the flow is steady, whereas once ${\sf m}>{\sf m}_c$ the flow shows an unsteady, time-periodic character. It must be emphasized that the latter occurs even though the driving mechanism is time-independent.  

The rigorous mathematical analysis of this type of bifurcation for {\em bounded} flow,  including stability properties of the bifurcating branch, has received a number of important contributions, beginning with the works  of Iudovich \cite{Yu}, Joseph \& Sattinger \cite{JS}, and Iooss \cite{Io} in the early 1970. In particular, these papers  laid the foundation for a rigorous understanding of complicated bifurcation phenomena occurring in the Taylor-Couette experiment; see \cite{CI}. 

However, it must also be emphasized that the approaches employed by these authors --mostly resembling  ideas  introduced by E.~Hopf in \cite{Hopf} on similar  problems for systems with a finite degree of freedom-- do not apply to the case of an {\em unbounded} flow. As a result, the important time-periodic bifurcation phenomenon occurring in the flow of a viscous liquid past body, like a  cylinder (in 2D) or a ball (in 3D), is left out. From a strictly technical viewpoint, this failure  is due to the circumstance that the above approaches require  the relevant time-independent, linearized operator, $\mathscr L$, to be continuously invertible in the appropriate Hilbert space where the problem is formulated. Now, while this condition is certainly satisfied if the region of flow is bounded, since in that case 0 can only be an eigenvalue for $\mathscr L$, in the case of an unbounded flow it fails, because 0 becomes a point of the essential spectrum \cite[Theorem 2 and Remark 2]{Bab}.  Nevertheless, as first pointed out and proved by Babenko \cite{Bab2}, the operator $\mathscr L$ becomes Fredholm of index 0 provided it is defined  in    the Banach space, $\mathcal B$, where steady-state solutions belong. Therefore, the bounded invertibility of $\mathscr L$, thus defined, is again ensured by requiring that 0 is not an eigenvalue.  In the light of these considerations, it becomes  natural to formulate the time-periodic bifurcation problem in the space $\mathcal B$, an approach first taken by Babenko \cite{Bab2}, and, successively extended and improved by Sazonov \cite{Saz}. 

However, this kind of procedure  has two drawbacks. On the one hand, it gives up the simplicity of the Hilbert-space formulation, and, on the other hand and more importantly, it is not able to cover the case of time-periodic bifurcation of plane flow past a cylinder \cite[p. 39]{Bab3}. Motivated  by the latter, in \cite{GaBi}  the present author has introduced  a different method for the study of time-periodic bifurcation of viscous flow that allows him to overcome both drawbacks. The method stems from the observation that, in the case of an unbounded flow, the (time-independent) time-average over a period, $v$, of the sought solution, and the ``purely periodic'' (time-dependent) component, $w$, belong, in general, to two {\em different} function spaces, with, in particular, $v\in \calb$.  With this in mind, the original time-dependent equation can be {\em equivalently} rewritten as {\em two  coupled equations}, one of the elliptic type (for $v$), and the other of parabolic type (for $w$). The problem then simplifies to a great extent, in that one can show that, in order to obtain the desired bifurcation result, it suffices to investigate, basically, only the properties of   the evolution equation which is proved to be naturally formulated in the {\em same Hilbert-space framework as that of bounded flow}. 

We believe that the method introduced in \cite{GaBi}  could be very useful in many other problems of mathematical physics, and, in particular, those regarding phenomena occurring  in unbounded spatial regions. 

For this reason, the main objective of this paper (Section 3) is to employ the basic ideas introduced in \cite{GaBi} to prove an  abstract time-periodic bifurcation result that could be applied to more general problems; see \theoref{3.1}.  As hinted earlier on, this theorem is formulated for the coupled systems constituted by a time-independent and a first order time-dependent equation in  Banach and Hilbert spaces, respectively; see \eqref{3.5}.\footnote{We wish to remark that our approach also admits of a straightforward extension to Banach spaces; see \remref{3.2}.}  Under suitable regularity conditions on the nonlinearities (see (H4) and \remref{3.3}) and technical assumptions (see (H3)), we then show the existence of a one-parameter family of bifurcating time-periodic solutions,  provided the spectrum of the relevant linearized operators satisfies certain specific conditions (see (H1), (H2), (H5)). Roughly speaking, they amount to assume that the linear (time-independent) operator involved in the evolution equation possesses a pair of simple, purely imaginary, complex conjugate eigenvalues, ``crossing'' the imaginary axis with non-zero speed; see also \remref{3.1}. Moreover, we show that this bifurcating branch is unique, and that the type of bifurcation can only be super- or sub-critical.   

The second part of the paper (Section 4) is dedicated to the application of \theoref{3.1} to the study of time-periodic bifurcation of a steady-state solution to the Navier-Stokes equation in an exterior three-dimensional domain (flow past a body). In particular, we show that all technical assumptions of \theoref{3.1} are indeed met (see \propref{4.1}--\propref{2.5}) so that the results stated in \theoref{3.1}, under the above mentioned hypotheses on the spectrum, apply. We wish to stress out that our results differ from those of \cite{Saz} on the one hand, because they are obtained, basically, in a Hilbert-space framework, and, on the other hand, because unlike \cite{Saz}, we also show the uniqueness property of bifurcating solutions.            
 
\section{Notation} The symbols $\nat$, $\mathbb Z$,  and ${\mathbb R}$, $\mathbb C$ stand, in the order,  for the sets of positive and relative integers, and the fields of real and complex numbers.

$\Omega$ denotes a fixed  exterior domain of $\real^3$, namely, the complement of the closure of a bounded, open,  and simply connected set, $\Omega_0\subset\mathbb R^3$. We shall assume  $\Omega$ of class $C^2$, and  take the origin $O$ of the coordinate system in $\Omega_0$.  Also, we denote by $R_*>0$ a number such that the closure of $\Omega_0$ is strictly contained in $\{\bfx\in\real^3: (x_1^2+x_2^2+x_3^2)^{\frac12}<R_*\}$.  

For $R\ge R_*$, we let
$$
\Omega_R=\Omega\cap \{\bfx\in\real^2: (x_1^2+x_2^2+x_3^2)^{\frac12}<R\}\,,\ \ \Omega^R=\Omega-\bar{\Omega_R}\,,
$$
where the bar denotes closure. 

We set $\bfu_t:=\partial \bfu/\partial t$, $\partial_1\bfu:=\partial \bfu/\partial x_1$, and indicate by $D^2\bfu$ the matrix of the second derivatives of $\bfu$.

For an open and connected set ${A} \subseteq {\mathbb R}^3,$ $L^q (A)$, $L^q_{loc}(A)$, $1\leq q \leq \infty,$  
$W^{m,q}({A}),$ $W_0^{m,q}(A)$, $m \geq 0,$  $(W^{0,q}\equiv W^{0,q}_0\equiv L^q$), stand for the usual Lebesgue and Sobolev classes, respectively, of real or complex functions.\Footnote{We shall use the same font style to denote scalar, vector and tensor
function spaces.}   
Norms in $L^q(A)$ and $W^{m,q}(A)$ are indicated by $\|.\|_{q,A}$ and $\|.\|_{m,q,A}$. The scalar product of functions $u,v\in L^2(A)$ will be denoted by $\langle u,v\rangle_A$. In the above notation,  the symbol $A$ will be omitted, unless confusion arises. 

As customary, for $q\in [1,\infty]$ we let $q'=q/(q-1)$ be its H\"older conjugate.  

By $D^{1,q}(\Omega)$, $1<q<\infty$, we denote the space of (equivalence classes of) functions $u$ such that
$ 
\|\nabla u\|_q<\infty\,.
$ 
Moreover, setting,
$$
\cald(\Omega):=\{\bfu\in C_0^\infty(\Omega):\Div\bfu=0\}
$$ 
we let $\mathcal D_0^{1,2}(\Omega)$ be the completion of $\cald(\Omega)$ in the norm $\|\nabla (\cdot)\|_2$, and set
$$
Z^{2,2}(\Omega):=W^{2,2}(\Omega)\cap \cald_0^{1,2}(\Omega)\,.
$$

Furthermore, we denote by $H_q(\Omega)$, $1<q<\infty$, ($H_2(\Omega)\equiv H(\Omega)$) the completion of $\cald(\Omega)$ in the norm $L^q(\Omega)$  and let  
 ${\rm P}_q$ be the (Helmholtz)  projection from $L^q(\Omega)$ onto $H_q(\Omega)$. ${\rm P}_q$ is independent of $q$ \cite[\S III.1]{GaB}, so that we shall simply denote it by ${\rm P}$.

We define
$$
X^{2,\frac43}(\Omega):=\big\{\bfu 
: \bfu\in L^{4}(\Omega)\cap D^{1,2}(\Omega)\cap D^{1,\frac{12}{5}}(\Omega) , \partial_1\bfu, D^2\bfu \in L^{\frac43}(\Omega)\big\}
$$
and 
$$
X^{2,\frac43}_0(\Omega):=\Big\{\bfu\in X^{2,\frac43}(\Omega): \Div\bfu=0\,,\ \bfu|_{\partial\Omega}=\0\Big\}\,.
$$
As is known, $X^{2,q}(\Omega)$ and $X^{2,q}_0(\Omega)$ become Banach spaces when endowed
with the ``natural'' norm
$$
\|\bfu\|_{X^{2,\frac43}}:=\|\bfu\|_{4}+\|\nabla \bfu\|_2+\|\nabla \bfu\|_{\frac{12}{5}}+\|\partial_1\bfu\|_{\frac43}+\|D^2\bfu\|_{\frac43}\,;
$$
see \cite{GaRb}.
\Br A function $\bfu\in X^{2,\frac43}(\Omega)$ decays to $0$ as $|x|\to\infty$ in a well defined sense. Precisely
$$
\lim_{R\to\infty}\int_{S_2}|\bfu(R,\Theta)|^{\frac {12}{5}}d\Theta=0
$$
where $S_2$ is the unit sphere in $\real^3$; see \cite[Lemma II.6.3]{GaB}.
\ER{2.1}

If $M$ is a map between two spaces, we denote by ${\sf D}\,[M]$, ${\sf N}\,[M]$ and ${\sf R}\,[M]$ its domain, null space and range, respectively.

In the following, $B$ is a real Banach space with associated norm $\|\cdot\|_B$.

By $B_{\mathbb C}:=B+{\rm i}\, B$ we denote the complexification of $B$.

For $q\in [1,\infty]$,  $L^q(-\pi,\pi;B)$ is the space of functions
$u:(-\pi,\pi)\rightarrow B$ such that 
$$
\left( \Int{\pi}{\pi}\|u(t)\|_B^q \right)^{\frac 1q}<\infty, \ \ \mbox{if 
$q\in [1,\infty)\,;$}\ \ \  
\essup{t\in[-\pi,\pi]}\|u(t)\|_B <\infty, \ \ \mbox{if $q=\infty.$}
$$

Given a function $u\in L^1(-\pi,\pi;B)$,
we let $\bar u$ be its average over $[-\pi,\pi]$, namely,
$$
{\bar u}:=\Frac{1}{2\pi}\int_{-\pi}^{\pi}u(t)dt\,.
$$
Furthermore, we shall say that
$u$ is {\em $2\pi$-periodic}, if $u(t+2\pi)=u(t)$, for a.a. $t\in \real$.
We then define
$$\ba{rl}\smallskip
\mathscr W^{2}_{2\pi,0}(\Omega):=\Big\{\bfu\in
 L^2(-\pi,\pi;Z^{2,2}(\Omega))\ \mbox{and}\ &\!\!\!\! \bfu_t\in L^{2}(-\pi,\pi;H(\Omega)):\\ &\!\!\!\!\bfu \ \mbox{is $2\pi$-periodic with $\bar\bfu=\0$}\Big\} 
\ea$$ 
with associated norm
$$
\|u\|_{\mathscr W^{2}_{2\pi,0}}:=\left(\int_{-\pi}^{\pi}\|\bfu_t(t)\|_{2}^2dt\right)^{1/2}+\left(\int_{-\pi}^{\pi}\|\bfu(t)\|_{2,2}^2dt\right)^{1/2}\,.
$$
\Br
Since $W^{2,2}\subset W^{1,6}$, from \cite[Theorem II.9.1]{GaB} it follows that if $\bfw\in\mathscr W_{2\pi,0}^2(\Omega)$ then
$$
\lim_{|x|\to\infty}|\bfw(x,t)|=0\, \ \mbox{uniformly in $x$, for a.a. $t\in [-\pi,\pi]$.}
$$
\ER{2.2}

Setting $$\Omega_{2\pi}:=\Omega\times [-\pi,\pi]$$
we define
$$
\mathscr L_{2\pi,0}(\Omega):=\Big\{\bfu\in
L^2(\Omega_{2\pi})):\ \bfu \ \mbox{is $2\pi$-periodic with $\bar\bfu=\0$}\Big\}\,, 
$$
and its subspace
$$
\mathscr H_{2\pi,0}(\Omega):=\Big\{\bfu\in
L^2(-\pi,{\pi}; H(\Omega)):\ \bfu \ \mbox{is $2\pi$-periodic with $\bar\bfu=\0$}\Big\}\,. 
$$
Moreover, for $\bfu,\bfv\in \mathscr L^2_{2\pi,0}(\Omega)$ we put
$$
(\bfu|\bfv):=\int_{-\pi}^{\pi}\langle\bfu(t),\bfv(t)\rangle\,dt\,.
$$

Finally, by  $c$, $c_0$, $c_1$, etc.,  we denote positive constants, whose particular value is unessential to the context. When we wish to emphasize
the dependence of $c$ on some parameter $\xi$, we shall write  $c(\xi)$.

\setcounter{equation}{0}
\section{An Abstract Bifurcation Theorem}
Objective of this section is to prove a  time-periodic bifurcation result for a general class of  equations in Banach spaces. Before proceeding in that direction, however, we first would like to make some comments that will also provide the motivation of our approach.

Many evolution problems in mathematical physics can be formally written in the form
\be
u_t+L(u)=N(u,\mu)\,,
\eeq{3.1}
where $L$ is a linear differential operator (with appropriate {\em homogeneous} boundary conditions), and $N$ is a nonlinear operator depending on the parameter $\mu\in\real$, such that $N(0,\mu)=0$ for all admissible values of $\mu$. Then, roughly speaking, time-periodic bifurcation for  \eqref{3.1} amounts to show the existence a family of non-trivial time-periodic solutions $u=u(\mu;t)$ of (unknown) period $T=T(\mu)$ ($T$-{\em periodic} solutions) in a neighborhood of $\mu=0$, and such that $u(\mu;\cdot)\to 0$ as $\mu\to 0$. Setting $\tau:=2\pi\,t/T\equiv \omega\, t$, \eqref{3.1} becomes
\be 
\omega\,u_\tau+L(u)=N(u,\mu)
\eeq{3.2}
and the problem reduces to find a family of $2\pi$-periodic solutions to \eqref{3.2} with the above properties. We now write $u=\bar{u}+(u-\bar{u}):=v+w$ and observe that \eqref{3.2} is formally equivalent to the following two equations
\be\ba{ll}\medskip 
L(v)=\bar{N(v+w,\mu)}:=N_1(v,w,\mu)\,,\\ \omega\,w_\tau+L(w)=N(v+w,\mu)-\bar{N(v+w,\mu)}:=N_2(v,w,\mu)\,.\ea
\eeq{3.3}
At this point, the crucial issue is that in many applications --typically when the physical system evolves in an {\em unbounded spatial region}-- the ``steady-state component" $v$ lives in function spaces with quite less ``regularity''\footnote{Here `regularity' is meant in the sense of behavior at large spatial distances.} than the space where the ``purely periodic" component $w$ does. For this reason, it is much more appropriate to study the two equations in \eqref{3.3} in two {\em different} function classes. As a consequence, even though {\em formally}  being the same as differential operators, the operator $L$ in \eqref{3.3}$_1$ acts on and ranges into spaces  different than those the operator $L$ in \eqref{3.3}$_2$ does. With this in mind, \eqref{3.3} becomes
$$
L_1(v)=N_1(v,w,\mu)\,;\ \ \omega\,w_\tau+L_2(w)=N_2(v,w,\mu)\,.
$$   
\par
The general abstract theory that we are about to describe stems exactly from the above considerations. 
  
To this end, let $\mathcal X, \mathcal Y$,  be  Banach spaces with norms $\|\cdot\|_{\mathcal X}$, $\|\cdot\|_{\mathcal Y}$, respectively, and let  $\mathcal H$ be a Hilbert space with norm $\|\cdot\|_{\mathcal H}$ and corresponding scalar product $\langle\cdot,\cdot\rangle$.\footnote{Without any risk of confusion, we use here the same symbol as the $L^2$-scalar product introduced earlier on.} Moreover,  denote by 
$$
L_1:\mathcal X\mapsto \mathcal Y\,,
$$
a bounded linear operator, and by
$$
L_2:{\sf D}\,[L_2]\subset \mathcal H\mapsto \mathcal H\,,
$$
a densely defined, closed linear operator, with a non-empty resolvent set ${\sf P}(L_2)$. For a fixed (once and for all) $\theta\in {\sf P}(L_2)$ we denote by $\mathcal W$ the linear subspace of $\mathcal H$ closed under the norm $\|w\|_{\mathcal W}:=\|(L_2+\theta\,I)w\|_{\mathcal H}$, where $I$ stands for the identity operator. We then define the following spaces
$$\ba{rl}\medskip
\mathcal H_{2\pi,0}&\!\!\!:=\big\{w\in L^2(-\pi,\pi;\mathcal H): \mbox{$2\pi$-periodic with} \ \bar{w}=0\big\} 
\\\medskip\calw_{2\pi,0}&\!\!\!:=\big\{w\in L^2(-\pi,\pi;\calw)\,, \  w_t\in L^{2}(-\pi,\pi;\mathcal H): \mbox{$2\pi$-periodic with} \ \bar{w}=0\big\}\,,
\ea$$ 
with corresponding norms
$$\ba{ll}\medskip
\|w\|_{\calh_{2\pi,0}}:=\Big(\Int{-\pi}\pi\|w(s)\|_{\calh}^2ds\Big)^{\frac12}
\\
\|w\|_{\calw_{2\pi,0}}:=\Big(\Int{-\pi}\pi\left(\|w(s)\|_{\calw}^2+\|w_s(s)\|^2_{\calh}\right)ds\Big)^{\frac12}\,.
\ea
$$ 
The scalar product in $\calh_{2\pi,0}$ is defined by\footnote{Without any risk of confusion, we use here the same symbol as the $\mathscr H_{2\pi,0}$-scalar product introduced earlier on.}
$$
(w_1|w_2):=\int_{-\pi}^{\pi}\langle w_1(s),w_2(s)\rangle\,ds\,.
$$
Next, let
$$
N: \calx\times \calw_{2\pi,0}\times \real\mapsto \caly\oplus \calh_{2\pi,0}   
$$
be a (nonlinear) map satisfying the following properties:
\be\ba{rl}\medskip
N_1&\!\!\!: (v,w,\mu)\in\calx\times \calw_{2\pi,0}\times \real\mapsto \bar{N(v,w,\mu)}\in \caly
\\
N_2&\!\!\!:=N-N_1:\calx\times \calw_{2\pi,0}\times \real\mapsto \calh_{2\pi,0}\,.
\ea
\eeq{3.4} 
We can then formulated the following.\smallskip\par 
Bifurcation Problem: {\em Find a neighborhood of the origin $U(0,0,0)\subset \calx\times \calw_{2\pi,0}\times \real$ such that the equations
\be
L_1(v)=N_1(v,w,\mu)\,,\ \mbox{in $\caly$}\,;\ \ \omega\, w_\tau +L_2(w)=N_2(v,w,\mu)\,,\ \mbox{in $\calh_{2\pi,0}$}\,,
\eeq{3.5}
possess there a family of non-trivial $2\pi$-periodic solutions $(v(\mu),w(\mu;\tau))$ for some $\omega=\omega(\mu)>0$,  such that $(v(\mu),w(\mu;\cdot))\to 0$ in $\calx\times\calw_{2\pi,0}$ as $\mu\to0$.}\smallskip\par
Whenever the Bifurcation Problem admits a positive answer, we say that  $(u=0,\mu=0)$ is a {\em bifurcation point}. Moreover, the bifurcation is called {\em supercritical} [resp. {\em subcritical}] if the family of solutions $(v(\mu),w(\mu;\tau))$ exists only for $\mu>0$ [resp. $\mu<0$]. 
\medskip\par
With a view to solve the above problem, we begin to make the following  assumptions (H1)--(H5) on the involved operators.
\begin{itemize}
  \item[(H1)] $L_1$ is a homeomorphism\,;
  \item[(H2)] The spectrum $\sigma(L_2)$ (computed with respect to $\calh_{\mathbb C}$) contains a simple eigenvalue $\nu_0:={\rm i}\,\omega_0$, $\omega_0>0$,\footnote{That is, ${\sf N}_{\mathbb C}[L_2-\nu_0I]\cap {\sf R}_{\mathbb C}[L_2-\nu_0I]=\{0\}$.} whereas $k\,\nu_0\not\in \sigma(L_2)$, for all $k\in\nat-\{0,1\}$\,;
  \item[(H3)] The operator
$$
\mathscr Q:w\in \calw_{2\pi,0}\mapsto \omega_0\,w_\tau+L_2(w)\in \calh_{2\pi,0}\,,
$$  
is Fredholm of index 0\,;
\item[(H4)] The nonlinear operators $N_1,N_2$
are analytic in the neighborhood $U_1(0,0,0)\subset \calx\times \calw_{2\pi,0}\times \real$, namely, there exists $\delta>0$ such that for all $(v,w,\mu)$ with $\|v\|_{\calx}+\|w\|_{\calw_{2\pi,0}}+|\mu|<\delta$, the Taylor series 
$$\ba{ll}\medskip
N_1(v,w,\mu)=\Sum{k,l,m=0}{\infty}R_{klm}v^kw^l\mu^m\,,\\
N_2(v,w,\mu)=\Sum{k,l,m=0}{\infty}S_{klm}v^kw^l\mu^m\,,
\ea
$$
are absolutely convergent in $\caly$ and $\calh_{2\pi,0}$, respectively, for all $(v,w,\mu)\in U_1$. Moreover, we assume that the multi-linear operators $R_{klm}$ and $S_{klm}$ satisfy $R_{klm}=S_{klm}=0$ whenever $k+l+m\le1$, and $R_{011}=R_{00m}=S_{00m}=0$, all $m\ge2$.
\end{itemize}
\par
In order to prove our main \theoref{3.1}, we begin to draw a number of consequences from the above assumptions. In this regard,  let $v_0$ be the (unique) normalized eigenvector of $L_2$ corresponding to the eigenvalue $\nu_0$, and set
$$
v_1:=\Re[v_0\,{\rm e}^{-{\rm i}\,\tau}]\,,\ \ v_2:=\Im[v_0\,{\rm e}^{-{\rm i}\,\tau}]\,. 
$$    
\Bl Under the assumption {\rm (H2)}, we have ${\rm dim}\,{\sf N}\,[\mathscr Q]=2$, and $\{v_1,v_2\}$ is a basis in ${\sf N}\,[\mathscr Q]$.
\EL{3.1}
{\em Proof.} Clearly, $\mathcal S:= {\rm span}\,\{v_1,v_2\}\subseteq {\sf N}[\mathscr Q]$. Conversely, take $w\in {\sf N}[\mathscr Q]$, and expand it in Fourier series
$$
w=\sum_{\ell=-\infty}^\infty w_\ell\,{\rm e}^{-{\rm i}\,\ell\,\tau}\,;\ w_\ell:=\frac1{2\pi}\int_{-\pi}^\pi w(\tau)\,{\rm e}^{{\rm i}\,\ell\,\tau}\,d\tau\,,\ \ w_0\equiv\bar w=0.
$$
Obviously, $w_\ell\in \calw_{\mathbb C}\equiv {\sf D}_{\mathbb C}[L_2]$. From $\mathscr Q(w)=0$ we deduce 
$$
-\ell\,\mu_0\,w_\ell+L_2(w_\ell)=0\,,\ \ w_\ell\in{\sf D}_{\mathbb C}[L_2]\,,\ \ \ell\in\mathbb Z, 
$$ 
which, by (H2) and the fact that $w_0=0$, implies $w_\ell=0$ for all $\ell\in \mathbb Z-\{\pm 1\}$. Thus, recalling that $\mu_0$ is simple, we infer  $w\in\cals$ and the lemma follows.\QED
\smallskip\par
Denote by $L^*_2$ the adjoint of  $L_2$. Since $\nu_0$ is simple (by (H2)), from classical results on Fredholm operators (e.g. \cite[Section 8.4]{Z}), it follows that there exists at least one  element  $v_0^*\in{\sf N}_{\mathbb C}[L_2^*-\nu_0\,I]$ such that $\langle v_0^*,v_0\rangle\neq 0$. Without loss, we may take 
\be
\langle v_0^*,v_0\rangle=\pi^{-1}\,.
\eeq{3.6}
We then define
$$
v_1^*:=\Re[v_0^*\,{\rm e}^{{\rm i}\,\tau}]\,,\ \ v_2^*:=\Im[v_0^*\,{\rm e}^{{\rm i}\,\tau}]\,, 
$$
and set
$$
\hat{\calh}_{2\pi,0}=\big\{w\in {\calh}_{2\pi,0}: \ (w|v_1^*)=(w|v_2^*)=0\big\}\,,\ \  
\hat{\calw}_{2\pi,0}={\calw}_{2\pi,0}\cap \hat{\calh}_{2\pi,0}\,.
$$
For future reference, we observe that with the normalization \eqref{3.6}, it follows that
\be\ba{ll}\medskip 
(v_1|v_1^*)=(v_2|v_2^*)=1\,,\ \ (v_2|v_1^*)=(v_1|v_2^*)=0\,,\\
( (v_1)_\tau|v_1^*)=0\,,\ \ ((v_1)_\tau|v_2^*)=-1\,. 
\ea
\eeq{3.7}
\Bl Let {\rm (H2)} and {\rm (H3)} hold. Then, the operator $\mathscr Q$ maps  $\hat{\calw}_{2\pi,0}$ onto $\hat{\calh}_{2\pi,0}$ homeomorphically.
\EL{3.2}
{\em Proof.} By (H3), $\mathscr Q$ is Fredholm of index 0, whereas by \lemmref{3.1} ${\rm dim}\,{\sf N}\,[\mathscr Q]=2$. From  classical theory of Fredholm operators (e.g. \cite[Proposition 8.14(4)]{Z}) it then follows that ${\rm dim}\,{\sf N}\,[\mathscr Q^*]=2$ 
where
$$
\mathscr Q^*=\omega_0(\cdot)_\tau+L_2^*
$$
is the adjoint of $\mathscr Q$. In view of the stated properties of $v_0^*$, we infer that ${\rm span}\,\{v_1^*,v_2^*\}={\sf N}\,[\mathscr Q^*]$, and the lemma follows from another classical result on Fredholm operators (e.g. \cite[Proposition 8.14(2)]{Z}).\QED
\smallskip\par
With this result in hand, we shall now follow a more or less standard procedure to show that our Bifurcation Problem has in fact a solution. To this end,  in order to ensure the the solutions we are looking for are non-trivial, we endow \eqref{3.5} with the side condition 
\be  
(w|v_1^*)=\varepsilon\,,\ \ (w|v_1^*)=0\,, 
\eeq{3.8}
where $\varepsilon$ is a real parameter ranging in a neighborhood of $0$. 
\smallskip\par
We may then prove the main result of this section.
\Bt Suppose  {\rm (H1)--(H5)} hold and, in addition\tag{H6}
\be
(S_{011}(v_1)|v_1^*)\neq 0\ \  \,.
\eeq{3.9}
Then, the following properties are valid. \smallskip\\
{\rm (a)} {\rm Existence.} There are analytic families\renewcommand{\theequation}{\arabic{section}.\arabic{equation}}\setcounter{equation}{8}
\be
\big(v(\varepsilon),w(\varepsilon),\omega(\varepsilon),\mu(\varepsilon)\big)\in \calx\times \calw_{2\pi,0}\times \real_+\times\real
\eeq{fam}
satisfying \eqref{3.5}, \eqref{3.8}, for all $\varepsilon$ in a neighborhood $\mathcal I(0)$ and such that
\be
\big(v(\varepsilon),w(\varepsilon)-\varepsilon\,v_1,\omega(\varepsilon),\mu(\varepsilon)\big)\to (0,0,\omega_0,0)\ \ \mbox{as $\varepsilon\to 0$}\,.
\eeq{3.10}
\par\noindent
{\rm (a)} {\rm Uniqueness.}
There is a neighborhood  $$U(0,0,\omega_0,0)\subset \calx\times \calw_{2\pi,0}\times \real_+\times \real$$ such that every (nontrivial) $2\pi$-periodic solution to \eqref{3.5}, $(z,s)$, lying in $U$ must coincide, up to a phase shift, with that member of the family \eqref{fam} having $\varepsilon\equiv(s|v_1^*)$.
\smallskip\par\noindent
{\rm (a)} {\rm Parity.}  The functions $\omega(\varepsilon)$ and $\mu(\varepsilon)$ are even:
$$
\omega(\varepsilon)=\omega(-\varepsilon)\,,\ \ \mu(\varepsilon)=\mu(-\varepsilon)\,,\ \ \mbox{for all $\varepsilon\in\cali(0)$\,.} 
$$
Consequently, the bifurcation due to these solutions is either subcritical or supercritical, a two-sided bifurcation being excluded.\footnote{Unless $\mu\equiv 0$.}
\ET{3.1}
{\em Proof.} We scale $v$ and $w$  by setting
$v=\varepsilon\,{\sf v}$, ${w}=\varepsilon\, {\sf w}$, so that problem \eqref{3.5},
 \eqref{3.8} becomes
\be\ba{ll}\medskip  
L_1({\sf v})={\mathcal N}_1(\varepsilon, {\sf v},{\sf w},\mu)\,,\ \mbox{in $\caly$}\,;\\ 
\omega_0\, {\sf w}_\tau +L_2({\sf w)={\mathcal N}_2(\varepsilon, \omega, {\sf v},{\sf w},\mu)\,,\ \mbox{in $\calh_{2\pi,0}$}}\,,
\ \
({\sf w}|v_1^*)=1\,,\ \ ({\sf w}|v_1^*)=0\,,
\ea
\eeq{3.11}
where 
$$\ba{ll}\medskip
\mathcal N_1(\varepsilon,\sfv,\sfw,\mu):=(1/\varepsilon)\,N_1(\varepsilon{\sf v}, \varepsilon\sfw,\mu)\,, \\
\mathcal N_2(\varepsilon,\omega,\sfv,\sfw,\mu):=(1/\varepsilon)\,N_2(\varepsilon{\sf v}, \varepsilon\sfw,\mu)+(\omega_0-\omega){\sf w}_\tau\,.\ea
$$ Define the map
$$\ba{cc}\medskip
F: (\varepsilon, {\sf U}):=(\varepsilon,\mu, \omega,{\sf v},{\sf w})\in \cali(0)\times U(0)\times V(\omega_0)\times \calx\times \calw_{2\pi,0}\\ \smallskip
\mapsto
\Big( L_1(\sfv)-\mathcal N_1(\varepsilon,\sfv,\sfw,\mu),\ 
\mathscr Q(\sfw)-\mathcal N_2(\varepsilon,\omega,\sfv,\sfw,\mu),\
({\sf w}|v_1^*)-1,\ ({\sf w}|v_2^*)\Big)\\
\in \caly\times \calh_{2\pi,0}\times \real^2\,,
\ea
$$
with $U(0)$ and $V(\omega_0)$  neighborhoods of $0$ and $\omega_0$. Since, by (H4), we have in particular $\mathcal N_1(0,0,v_1,0)=
\mathcal N_2(0,\omega_0,v_1,0)=0$, using  
\eqref{3.7}$_1$ and \lemmref{3.1} we deduce that, at $\varepsilon=0$,  
the equation $F(\varepsilon,{\sf U})=0$ has the solution ${\sf U}_0=(0,\omega_0,0,v_1)$. Therefore, since by (H4) we have that $F$ is analytic at $(0,{\sf U_0})$, by the  analytic version of the Implicit Function Theorem (e.g. \cite[Proposition 8.11]{Z}), to show the existence statement -including the validity of \eqref{3.10}- it suffices to show that the Fr\'echet derivative, $DF(0,{\sf U}_0)$,  of $F$ with respect to ${\sf U}$ evaluated at $(0,{\sf U}_0$) is a bijection. 
Now,   in view of the assumption (H4), it easy to see that the Fr\'echet derivative of $\mathcal N_1$ at $(\varepsilon=0, \sfv=0,\sfw=v_1,\mu=0)$ is equal to 0, while that of $\mathcal N_2$ at $(\varepsilon=0, \omega=\omega_0,\sfv=0,\sfw=v_1,\mu=0)$ is equal to $ -\omega\,(v_{1})_\tau+\mu\,S_{011}(v_1)$\,.
Therefore, $DF(0,{\sf U}_0)$ is a bijection  if  we  prove that for any
$({\sf f}_1,{\sf f}_2,{\sf f}_3,{\sf f}_4)\in \caly\times\calh_{2,\pi,0}\times\real\times\real$, the following set of equations has one and only one solution $(\mu,\omega,\sfv,\sfw)\in \real\times\real\times \calx\times \calw_{2\pi,0}$:
\be\ba{rl}\medskip
{L}_1(\sfv)=&\!\!\!\!{\sf f}_1\ \ \mbox{in $\caly$}\\ \medskip
\mathscr Q(\sfw)= &\!\!\!\!-\omega\, (v_{1})_\tau+\mu\,S_{011}(v_1)+{\sf f}_2\ \ \mbox{in $\calh_{2\pi,0}$}\,,\\
({\sf w}|\bfv_1^*)=&\!\!\!\!{\sf f}_3\,,\ \ ({\sf w}|\bfv_2^*)={\sf f}_4 \ \ \mbox{in $\real$}\,,
\ea
\eeq{3.12}
In view of (H1), for any given ${\sf f}_1\in \caly$, equation \eqref{3.12}$_1$ has one and only one solution $\sfv\in \calx$. Therefore, it remains to prove the existence and uniqueness property only for the system of equations \eqref{3.12}$_{2-4}$
To this aim, we observe that, by \lemmref{3.2}, for a given ${\sf f}_2\in \calh_{2\pi,0}$,  equation \eqref{3.12}$_2$ possesses a unique solution $\sfw_1\in\hat{\calw}_{2\pi,0}$ if and only if its right-hand side is in $\hat{\calh}_{2\pi,0}$, namely,
$$
\big(-\omega\, (v_{1})_\tau+\mu\,S_{011}(v_1)+{\sf f}_2|v_1^*\big)=\big(-\omega\,( v_{1})_\tau+\mu\,S_{011}(v_1)+{\sf f}_2|v_2^*\big)=0\,. 
$$
Taking into account \eqref{3.7}$_{2}$ the above conditions will be satisfied provided we can find $\mu$ and $\omega$  satisfying the following algebraic system
\be\ba{rl}\medskip
\mu(\,S_{011}(v_1)|v_1^*)&\!\!\!\!=-({\sf f}_2|v_1^*)\\
\omega+\mu\,(\,S_{011}(v_1)|v_2^*)&\!\!\!\!=-({\sf f}_2|v_2^*)\,.
\ea
\eeq{3.13}
However, by virtue of \eqref{3.9}, this system possesses a uniquely determined solution $(\mu,\omega)$, which ensures the existence of a unique solution ${\sf w}_1\in \hat{\calw}_{2\pi,0}$ to \eqref{3.12}$_2$ corresponding to the  selected values of $\mu$ and $\omega$. We now set
$$ 
\sfw:={\sf w}_1+\alpha\,v_1+\beta\,v_2\,,\ \ \alpha\,,\, \beta\in\real\,.
$$
Clearly, by \lemmref{3.1}, ${\sfw}$ is also a solution to \eqref{3.12}$_2$. We then 
choose $\alpha$ and $\beta$ in such a way that $\sfw$ satisfies both conditions \eqref{3.12}$_{3,4}$  for any given ${\sf f}_i\in\real$, $i=1,2$. This choice is made possible by virtue of \eqref{3.7}$_1$.
We have thus shown that $DF(0,{\sf U}_0)$ is surjective. To show that it is also injective,
set ${\sf f}_i=0$ in \eqref{3.12}$_{2-4}$. From \eqref{3.13} and \eqref{3.9} it then follows $\mu=\omega=0$ which in turn implies, by \eqref{3.12}$_2$ and  \lemmref{3.1}, $\sfw=\gamma_1\,v_1+\gamma_2\,v_2$, for some $\gamma_i\in\real$, $i=1,2$. Replacing this information back in \eqref{3.12}$_{3,4}$ with ${\sf f}_3={\sf f}_4=0$, and using \eqref{3.7}$_1$ we conclude $\gamma_1=\gamma_2=0$, which proves  the claimed injectivity property.
Thus, $DF(0,{\sf U}_0)$ is a bijection, and the proof of the existence statement in (a) is completed. 
We shall next show the uniqueness statement in (b) by adapting to the present case the  argument of \cite[Theorem 8.B]{Z}. 
Let $(z,s)\in \calx\times\calw_{2\pi,0}$  be a  $2\pi$-periodic solution to \eqref{3.5} with $\omega\equiv\tilde{\omega}$ and $\mu\equiv\tilde\mu$.
By the uniqueness property associated with  the implicit function theorem, the proof of the claimed uniqueness
amounts to show that we can find a sufficiently small $\rho>0$ such that if
\be
\|z\|_{\calx}+\|s\|_{\calw_{2\pi,0}}+|\tilde\omega-\omega_0|+|\tilde\mu|<\rho\,,
\eeq{3.14}
then there exists a neighborhood of $0$, $\cali(0)\subset\real$, such that
\be\ba{cc}\medskip
s=\eta\, v_1+\eta\,{\sf s}\,,\, \ z=\eta\,{\sf z}\,, \ \mbox{for all $\eta\in\cali(0)$},\, \\
|\tilde\omega-\omega_0|+|\tilde\mu|+\|{\sf z}\|_{\calx}+\|{\sf s}\|_{\calw_{2\pi,0}}\to 0\ \ \mbox{as $\eta\to 0$}\,.\ea
\eeq{3.15}
To this end, we notice that, by \eqref{3.7}$_1$, we may write
\be
s={\sigma} +\ts
\eeq{3.16}
where ${\sigma}=(s|v^*_1)\,v_1+(s|v^*_2)\,v_2$ and 
\be
(\tilde{\sf s}|v^*_i)=0\,,\ \ i=1,2\,.
\eeq{3.17}
We next make the simple but important observation that if we modify $s$ by a  constant phase shift in time, $\delta$, namely, $s(\tau)\to s(\tau+\delta)$,  the shifted function is still  a $2\pi$-periodic solution to \eqref{3.5}$_2$ and, moreover, by an appropriate choice of $\delta$, 
\be   
{\sigma}=\eta\, v_1\,,
\eeq{3.18}
with $\eta=\eta(\delta)\in\real$. (The proof of \eqref{3.18} is straightforward, once we take into account the definition of $v_1$ and $v_2$.)
Notice that from \eqref{3.14}, \eqref{3.16}--\eqref{3.18} it follows that
\be
|\eta| +
\|{\ts}\|_{\calw_{2\pi,0}}
\to 0 \ \ \mbox{as $\rho\to 0$}\,.
\eeq{3.19} 
From \eqref{3.5} we thus get 
\be 
{L}_1(z)= N_1(z,\eta\, v_1+\ts,\tilde\mu) 
\eeq{3.20}
and, recalling \lemmref{3.1}, 
\be
\mathscr Q(\ts)=\eta(\omega_0-\omega)(v_1)_\tau+(\omega_0-\omega)\ts_\tau+N_2(z,\eta\,v_1+\ts,\tilde\mu)\,.
\eeq{3.21}
In view of (H4) and \eqref{3.14}, we easily deduce deduce
$$
N_1(z,\eta\, v_1+\ts,\tilde\mu)=R_{110}z(\eta\,v_1+\ts)+R_{101}z\tilde\mu+R_{020}(\eta\,v_1+\ts)^2+n_1(z,\eta,\ts,\tilde\mu)\,,
$$
where 
$$
\|n_1(z,\eta,\ts,\tilde\mu)\|_{\caly}\le \epsilon(\rho) \,\left(\|z\|_{\calx}+\|\ts\|_{\calw_{2\pi,0}}+\eta^2\right)\,,\ \ \
\epsilon (\rho)\to 0\ \mbox{as}\ \rho\to 0\,,$$
so that, by \eqref{3.20} and (H1) we obtain by taking $\rho$ sufficiently small
\be
\|z\|_{\calx}\le c_1\,\big(|\eta|^2+\|\ts\|_{\calw_{2\pi,0}}^2+\epsilon(\rho)\|\ts\|_{\calw_{2\pi,0}}\big)\,.
\eeq{3.22}
Likewise,
\be\ba{rl}\medskip
N_2(z,\eta\,v_1+\ts,\tilde\mu)
=&\!\!\!\!S_{011}(\eta\,v_1+\ts)\tilde\mu+S_{110}z(\eta\,v_1+\ts)+S_{101}z\tilde\mu\\&+S_{200}z^2+S_{020}(\eta\,v_1+\ts)^2+n_2(z,\eta,\ts,\tilde\mu)\,,
\ea
\eeq{3.23}
where $n_2$ enjoys the same property as $n_1$.
From \eqref{3.21},  \eqref{3.23} and \eqref{3.7}$_1$ we infer, according to \lemmref{3.2}, that the
following (compatibility) conditions must be satisfied
$$\ba{ll}\medskip
-\eta\,\tilde\mu\,(S_{011}(v_1)|v_1^*)=\big((\omega_0-\omega)\ts_\tau+ S_{011}\ts\tilde\mu+S_{110}z(\eta\,v_1+\ts)|v_1^*\big)\\ \medskip
\hspace*{3.5cm}+\big(S_{200}z^2+S_{020}(\eta\,v_1+\ts)^2|v_1^*\big)+(n_2|v_1^*)\\ \medskip
\eta\,(\omega-\omega_0)=\big((\omega_0-\omega)\ts_\tau+ S_{011}\ts\tilde\mu+S_{110}z(\eta\,v_1+\ts)|v_2^*\big)\\ \medskip
\hspace*{3.5cm}+\big(S_{200}z^2+S_{020}(\eta\,v_2+\ts)^2|v_2^*\big)++(n_2|v_2^*)\,,
\ea
$$
so that, from \eqref{3.9} and the property of $n_2$  we show
\be \ba{ll}\medskip
|\eta|\,\big(|\tilde\mu|+|\omega-\omega_0|\big)\le c_2 \big(|\omega-\omega_0|+|\tilde\mu|\big)\,\|\ts\|_{\calw_{2\pi,0}}
+|\eta|\,\|z\|_{\calx}+\|z\|_{\calx}^2\\
\hspace*{3.5cm}+\|\ts\|_{\calw_{2\pi,0}}^2+\eta^2\big)+\epsilon(\rho)\big(\|z\|_{\calh}+\|\ts\|_{\calw_{2\pi,0}}\big)\,.\ea
\eeq{3.24}
Also, applying \lemmref{3.2} to \eqref{3.21} and using \eqref{3.23}, \eqref{3.14} with $\rho$ sufficiently small we get
\be
\|\ts\|_{\calw_{2\pi,0}}\le c_3\,\big(|\eta|\,(|\tilde\mu|+|\omega-\omega_0|)+(|\eta|+|\tilde\mu|+\epsilon(\rho))\,\|z\|_{\calx}+\|z\|_{\calx}^2+\eta^2\big)\,.
\eeq{3.25}
Summing side by side \eqref{3.22}, \eqref{3.24} and $(1/(2c_3))\times$\eqref{3.25}, and taking again $\rho$ small enough, we thus arrive at
$$
|\eta|\,\big(|\tilde\mu|+|\omega-\omega_0|\big)+\|z\|_{\calx}+\|\ts\|_{\calw_{2\pi,0}}\le c_4\,\eta^2\,,
$$
from which we establish the validity of \eqref{3.15}$_2$, thus concluding the proof of the uniqueness property (b). Finally, in order to show the parity property in (c), we notice that if $\big(v(-\varepsilon),w(-\varepsilon;\tau)\big)$ is the solution corresponding to $-\varepsilon$, we have $\big(w(-\varepsilon;\tau+\pi)|v_1^*\big)=\varepsilon \,v_1$, which, by part (b), implies that, up to a phase shift, $\big(v(-\varepsilon),w(-\varepsilon;\tau)\big)=\big(v(\varepsilon),w(\varepsilon;\tau)\big)$. This, in turn, furnishes $\omega(-\varepsilon)=\omega(\varepsilon)$ and $\mu(-\varepsilon)=\mu(\varepsilon)$. From the latter and the analyticity of $\mu$ we then obtain
that either $\mu\equiv0$ or else there is an integer $k\ge 1$ such that 
$$
\mu(\varepsilon)=\varepsilon^{2k}\mu_k+O(\varepsilon^{2k+2})
\ \ \mu_k
\in\real-\{0\}\,.
$$
Thus, $\mu(\varepsilon)<0$ or $\mu(\varepsilon)>0$, according to whether $\mu_k$ is negative or positive. The theorem is completely proved.     
\QED
\Br By means of a classical result on eigenvalues perturbations, we can give an equivalent (and more familiar) formulation of \eqref{3.9}. To this end, let
$$
L_2(\mu):=L_2+\mu\,S_{011}\,,
$$
and observe that, by (H2), $\nu_0$ is a simple eigenvalue of $L_2(0)\equiv L_2$. Therefore, denoting by $\nu(\mu)$ the eigenvalues of $L_2(\mu)$, we know (e.g.  \cite[Proposition 79.15 and Corollary 79.16]{Z1}) that in a neighborhood of $\mu=0$ the map $\mu\mapsto\nu(\mu)$ is well defined and of class $C^\infty$, and that
$$
\nu'(0)=\langle v_0^*, S_{011}(v_0)\rangle\,.
$$
With the help of the latter and  a straightforward calculation we then show that \eqref{3.9} is {\em equivalent} to the condition
$$
\Re[\nu'(0)]\neq 0\,,
$$
which in turn tells us that the  eigenvalue $\nu(\mu)$ must cross the imaginary axes with ``non-zero speed''. \ER{3.1}
\Br The arguments used in the proof of \theoref{3.1} go through in the more general case where the evolution equation \eqref{3.5}$_2$ is formulated in a Banach space, provided we modify (H3) by adding the assumption that ${\sf N}\,[\mathscr Q]$ is two-dimensional. However, we preferred the Hilbert formulation just to emphasize that, as shown in the next section, time-periodic bifurcation of a Navier-Stokes steady-state flow past an obstacle can be safely and successfully handled in the simpler Hilbert-space framework.   
\ER{3.2}
\Br The assumption of analyticity of $N_1$ and $N_2$ with respect to $(v,w,\mu)$ is not necessary. Actually, a suitably modified version of \theoref{3.1} continues to hold if the nonlinear terms are of class $C^k$ in all variables, for some $k\ge 2$. In such a case, the family of branching solutions of \theoref{3.1} will be of class $C^{k-1}$ in the parameter $\varepsilon$. 
\ER{3.3}
\setcounter{equation}{0}
\section{Time-periodic Bifurcation of Steady-State Solutions to the Navier-Stokes Equations Past an Obstacle}
In this section we will apply the general theory developed in the previous one to the study of time-periodic bifurcation from a steady-state flow of a Navier-Stokes liquid past a three-dimensional obstacle. To this end, assume that an obstacle, $\mathscr B$, of diameter $d$ is placed in the flow of a Navier-Stokes liquid having an upstream velocity $\bfv_\infty$. 
Then, the bifurcation problem amounts to study the following set of (dimensionless) equations  
\be\ba{cc}\medskip\left.\ba{cc}\medskip
{\bfV}_t+\lambda(\bfV-\bfe_1)\cdot\nabla\bfV=\Delta\bfV-\nabla P\\
\Div\bfV=0
\ea\right\}\ \ \mbox{in $\Omega\times \real$}\\
\bfV=\bfe_1\ \ \mbox{at $\partial\Omega\times \real$}\,,\ea
\eeq{4.1}
with the further condition
\be
\lim_{|x|\to\infty}\bfV(x,t)=\0\,, \ \ \mbox{ $t\in \real$}\,.
\eeq{4.2}
Here $\bfV$ and $P$ are velocity and pressure fields of the liquid, $\Omega$ is the  region of flow, namely, the entire three-dimensional space exterior to $\mathscr B$,  $\bfe_1$ is a unit vector   parallel to $\bfv_\infty$, and $\lambda:=|\bfv_\infty|/(\bar{\nu}\,d)$, with $\bar{\nu}$ kinematic viscosity of the liquid, is the {\em Reynolds number}. It will be shown (see \propref{4.1}) that, under suitable assumptions on $\lambda_0$, the above equations  possess a unique steady-state solution branch $(\bfu(\lambda),p(\lambda))$, with $\lambda$ in  a neighborhood $U(\lambda_0)$. 
Writing $\bfV=\bfv(x,t;\lambda)+\bfu(x;\lambda)$, $P={\sf p}(x,t;\lambda)+p(x;\lambda)$,  equations \eqref{4.1}--\eqref{4.2} become 
\be\ba{cc}\medskip\left.\ba{rl}\medskip
{\bfv}_t\!+\!\lambda\big[(\bfv-\bfe_1)\cdot\nabla\bfv +\bfu(\lambda)\cdot\nabla\bfv+\bfv\cdot\nabla\bfu(\lambda)\big]\!=&\!\!\!\!\Delta\bfv-\nabla {\sf p}\\
\Div\bfv=&\!\!\!\!0
\ea\!\right\}\, \ \mbox{in $\Omega\times \real$}\\
\bfv=\0\ \ \mbox{at $\partial\Omega\times\real$}\,,\ea
\eeq{4.3}
with
\be
\lim_{|x|\to\infty}\bfv(x,t)=\0\,, \ \ \mbox{ $t\in \real$}\,.
\eeq{4.4}
Our bifurcation problem consists then in finding sufficient conditions for the existence of a   non-trivial family of time-periodic solutions to \eqref{4.3}--\eqref{4.4}, $(\bfv(\lambda),{\sf p}(\lambda))$, $\lambda\in U(\lambda_0)$, of period $T=T(\lambda)$ (unknown as well), such that $(\bfv(t;\lambda),\nabla{\sf p}(t;\lambda))\to (\0,\0)$ as $\lambda\to\lambda_0$.    

We shall show that \eqref{4.3}--\eqref{4.4} can be put in the form \eqref{3.5}, for an appropriate choice of the involved operators and  function spaces, and that if conditions (H1), (H2) and \eqref{3.9} hold, then the bifurcation result of \theoref{3.1} applies. 
\smallskip\par
In this regard, for  $\bfu_0\in X^{2,\frac43}(\Omega)$ and $\lambda_0>0$ define the operator 
\be
\mathscr L_1: \bfv\in X^{2,\frac43}_0\mapsto{\rm P}\,\big[\Delta\bfv+\lambda_0(\partial_1\bfv-\bfu_0\cdot\nabla\bfv-\bfv\cdot\nabla\bfu_0)\big]\in H_{\frac43}(\Omega)\,.
\eeq{4.5}
By the properties of the $X$- and $H$-spaces and the H\"older inequality, we easily show that $\mathscr L_1$ is well-defined. 
The following result holds.
\Bp $\mathscr L_1$ is Fredholm of index 0. Moreover, assume that $(\bfu_0,p_0)\in X^{2,\frac43}\times D^{1,\frac43}$ is a steady-state solution to problem \eqref{4.1}--\eqref{4.2} with $\lambda=\lambda_0$,  namely, $(\bfu_0,p_0)$ solves
\be 
\ba{ll}\medskip
\left.\ba{rl}\medskip
\Delta\bfu+\lambda\,\partial_1\bfu&\!\!\!=\lambda\,\bfu\cdot\nabla\bfu +\nabla p\\
\Div\bfu&\!\!\!=0\ea\right\}\ \ \mbox{in $\Omega$}\\
\bfu=\bfe_1\ \ \mbox{at $\partial\Omega$}\,,\ \ \Lim{|x|\to\infty}\bfu(x)=\0\,,
\ea
\eeq{4.6}
corresponding to $\lambda=\lambda_0$. Then, if ${\sf N}[\mathscr L_1]=\{0\}$,
problem \eqref{4.6} has a solution that is (real) analytic at $\lambda=\lambda_0$. Precisely,
there is a neighborhood $U(\lambda_0)$ of $\lambda_0$ and a  solutions family to \eqref{4.6},   $(\bfu(\lambda),p(\lambda))\in X^{2,\frac43}(\Omega)\times D^{1,\frac43}(\Omega)$,  $\lambda\in U(\lambda_0)$, such that the series
$$
\bfu(\mu+\lambda_0)=\bfu_0+\sum_{k=1}^\infty \mu^k\bfu_k\,,\ \ p(\mu+\lambda_0)=p_0+\sum_{k=1}^\infty \mu^kp_k\,,\ \ \mu:=\lambda-\lambda_0
$$
are absolutely convergent in $X^{2,\frac43}(\Omega)$ and $D^{1,\frac43}(\Omega)$, respectively.  
\EP{4.1}
{\em Proof.} The Fredholm property is shown in \cite[Theorem 3.1]{GaRb}. Next, we notice that setting $\bf\tilde\bfu:=\bfu-\bfu_0$, $\phi:=p-p_0$, from  \eqref{4.6} we deduce that $(\tilde\bfu,\mu)$ satisfies
\be
\mathscr F(\tilde\bfu,\mu):=\mathscr L_1(\tilde\bfu)-\mathscr N(\tilde\bfu,\mu)=\0
\eeq{4.7}
where
$$
\mathscr N(\tilde\bfu,\mu):={\rm P}\,\big[-\mu\,(\partial_1\tilde\bfu-\bfu_0\cdot\nabla\tilde\bfu-\tilde\bfu\cdot\nabla\bfu_0)-(\mu+\lambda_0)(\bfu_0\cdot\nabla\tilde\bfu+\tilde\bfu\cdot\nabla\bfu_0)\big]\,. 
$$
By the H\"older inequality, we show at once that the bilinear form 
$$
(\bfu_1,\bfu_2)\in X^{2,\frac43}(\Omega)\times X^{2,\frac43}(\Omega)\mapsto \bfu_1\cdot\nabla\bfu_2\in L^{\frac43}(\Omega)\,,
$$
is continuous, and therefore the operator 
$\mathscr N:(\tilde\bfu,\mu)\in X_0^{2,\frac43}\times\real\mapsto \mathscr N\in H_{\frac43}$ is analytic at any $(\tilde\bfu,\mu)$, and so is $\mathscr F:(\tilde\bfu,\mu)\in X_0^{2,\frac43}\times\real\mapsto \mathscr L_1-\mathscr N\in H_{\frac43}$. Now, $\mathscr F(\0,0)=0$, and, being ${\sf N}\,[\mathscr L_1]=\{0\}$ by assumption, the Fr\'echet derivative $D_{\tilde\bfu}\mathscr F(\0,0)\equiv\mathscr L_1$ is a homeomorphism. As a consequence the lemma follows from the analytic version of the Implicit Function Theorem (e.g.  \cite[Proposition 8.11]{Z}).  
\QED

We now introduce the operator 
\be\ba{ll}\medskip
\mathscr L_2: \bfv\in {\sf D}[\mathscr L_2]\!\subset\! H(\Omega)\mapsto -{\rm P}\,\big[\Delta\bfv+\lambda_0(\partial_1\bfv-\bfu_0\cdot\nabla\bfv-\bfv\cdot\nabla\bfu_0)\big]\!\in\! H(\Omega)\,,\\ \hspace*{1.5cm} {\sf D}[\mathscr L_2]:=Z^{2,2}(\Omega)\,.\ea
\eeq{4.8}
Since $Z^{2,2}(\Omega)$ is  dense in $H(\Omega)$, $\mathscr L_2$ is densely defined. Moreover, with the help of H\"older inequality and the embedding $W^{2,2}\subset W^{1,4}\subset L^{12}$ it is easy to check that ${\sf R}\,[\mathscr L_2]\in H(\Omega)$, provided $\bfu_0\in X^{2,\frac43}(\Omega)$.\footnote{See also \eqref{4.12}, \eqref{4.13}.} 
Our main objective is to show  that the intersection of the spectrum $\sigma(\mathscr L_2)$ (computed with respect to $H_{\mathbb C}$)  with $\{{\rm i}\real-\{0\}\}$ is constituted  at most by a finite or countable number of eigenvalues with finite multiplicity (see \propref{4.1}).

The proof of this property requires some   preparatory results.
\Bl Let $\omega\in \real-\{0\}$. Then, for a given $\bff\in L^2_{\mathbb C}(\Omega)$ there is a unique corresponding $(\bfu,p)\in W_{\mathbb C}^{2,2}(\Omega)\times D^{1,2}_{\mathbb C}(\Omega)$ such that
\be 
\ba{cc}\medskip
\left.\ba{rl}\medskip
\Delta\bfu+\lambda_0\,\partial_1\bfu-\i\,\omega\,\bfu&\!\!\!=\bff +\nabla p\\
\Div\bfu&\!\!\!=0\ea\right\}\ \ \mbox{in $\Omega$}\,,\\
\bfu=\0\ \ \mbox{at $\partial\Omega$}\,.
\ea
\eeq{2.5}
Moreover, there are constants $c$ and $c_0$ depending only on $\Omega$, such that $(\bfu,p)$ satisfies the following inequality
\be
\|D^2\bfu\|_2+|\omega|^{\frac12}\|\nabla\bfu\|_2+|\omega|\|\bfu\|_2+\|\nabla p\|_2\le c\,\|\bff\|_2\,,\ \ |\omega|\ge \max\{\lambda_0^2,1\}\,.
\eeq{2.6}

\EL{2.1}
{\em Proof.} The proof  is entirely analogous to that of \cite[Lemma 4.1]{GaBi}) and will be thus omitted.
\QED

\Bl  The operator
$$
\mathscr K:\bfv\in Z^{2,2}(\Omega)\mapsto \bfu_0\cdot\nabla\bfv+\bfv\cdot\nabla\bfu_0\in L^2(\Omega)
$$
is compact.
\EL{2.2}
{\em Proof.}  
We begin to recall the embeddings
\be\ba{ll}\medskip
Z^{2,2}(\Omega)\subset W^{1,4}(\Omega)\subset L^{12}(\Omega)\,,\\ 
Z^{2,2}(\Omega)\subset W^{1,4}(\Omega_R)\,\subset L^{12}(\Omega_R)\,, 
 \ \   \mbox{compact, for all $R>R_*$}\,.
\ea
\eeq{4.11}
Let $\{\bfv_n\}\subset Z^{2,2}(\Omega)$ with $\|\bfv_n\|_{2,2}=1$, for all $n\in\nat$, and let $\bar\bfv\in Z^{2,2}(\Omega)$ be its weak limit. Without loss of generality, we may assume $\bar\bfv=\0$, which gives $\mathscr K(\bar\bfv)=\0$.
For any $R>R_*$  we show, by H\"older inequality and \eqref{4.11}$_1$, that
\be
\|\bfu_0\cdot\nabla\bfv_n\|_2\le
\|\bfu_0\|_{4}\|\nabla\bfv_n\|_{4,\Omega_R}+c_1\,\|\bfu_0\|_{4,\Omega^R}\|\bfv_n\|_{2,2}\\
\eeq{4.12}
Likewise, 
\be
\|\bfv_n\cdot\nabla\bfu_0\|_2\le
\|\nabla\bfu_0\|_{\frac{12}{5}}\|\bfv_n\|_{12,\Omega_R}+c_2\,\|\nabla\bfu_0\|_{\frac{12}{5},\Omega^R}\|\bfv_n\|_{2,2}\,.
\eeq{4.13}
As a result, since $\bfu_0\in X^{2,\frac43}(\Omega)$,  by  \eqref{4.11}$_2$--\eqref{4.13}, and taking $R$ arbitrarily large, we may conclude
$$
\lim_{n\to\infty}\|\mathscr K(\bfv_n)\|_2=0\,.
$$
which proves
the claimed compactness property of $\mathscr K$,  and completes the proof of the proposition.\QED

\Bl Let $\bfu_0\in X^{2,\frac43}(\Omega)$,   
and let $\omega\in\real-\{0\}$. Then,\footnote{By $I$ we mean the identity operator in $ H_{\mathbb C}$.} the operator
\be 
\mathscr L_{\omega}:=\mathscr L_2-\i\,\omega I\,,
\eeq{2.13}
is Fredholm of index 0.
\EL{2.3}
{\em Proof.}
$\mathscr L_\omega$ is (graph) closed. In fact, this follows from \cite[Theorem 1.11 in Chapter IV]{Kato}, since $\mathscr L_\omega=\mathscr L_1+\mathscr K$, where $\mathscr L_1$ is a homeomorphism (\lemmref{2.1}) and thus obviously closed,  whereas  by \lemmref{2.2}, $\mathscr K$ is $\mathscr L_1$-compact. These two combined properties also show that \eqref{2.13} is Fredholm of index 0 (e.g. \cite[Theorem XVII.4.3]{GoGoKa}). The lemma is  proved.\QED  

We are now in a position to show the first main result of this section.
\Bp Let $\bfu_0\in X^{2,\frac43}(\Omega)$. Then $\sigma(\mathscr L_2)\cap\big\{{\rm i}\,\real-\{0\}\big\}$ consists, at
most, of a finite or countable number of eigenvalues, each of which is isolated and of finite
(algebraic)  multiplicity, that can only accumulate at 0. 
\EP{4.2}
{\em Proof.} By \lemmref{2.3} we know that $\mathscr L_\omega:H_{\mathbb C}(\Omega)\mapsto H_{\mathbb C}(\Omega)$ is an (unbounded) Fredholm operator of index 0, for all $\omega\in \real-\{0\}$. Thus, in view of well-known results (e.g. \cite[Theorem XVII.2.1]{GoGoKa}), in order to prove the stated property it is enough to show that there is $\bar\omega>0$  such that for all $|\omega|>\bar\omega$, ${\sf N}\,[\mathscr L_\omega]=\{0\}$. Now, the equation $\mathscr L_\omega(\bfv)=\0$ is equivalent to the following problem
\be
\ba{cc}\medskip
\left.\ba{rl}\medskip
\Delta\bfv+\lambda_0\,\partial_1\bfv-\i\,\omega\,\bfv&\!\!\!=\lambda_0\left(\bfu_0\cdot\nabla\bfv+\bfv\cdot\nabla\bfu_0\right) +\nabla p\\
\Div\bfv&\!\!\!=0\ea\right\}\ \ \mbox{in $\Omega$}\,,\\
\bfv=\0\ \ \mbox{at $\partial\Omega$}\,,
\ea
\eeq{2.12}
with $(\bfv,p)\in Z_{\mathbb C}^{2,2}(\Omega)\times D_{\mathbb C}^{1,2}(\Omega)$. Using \lemmref{2.1} and \eqref{2.6} in problem \eqref{2.12}, with the help of H\"older inequality we get, in particular, for all $|\omega|\ge \max\{\lambda_0^2,1\}$,
$$\ba{rl}\medskip
\|D^2\bfv\|_2+|\omega|^{\frac12}\|\nabla\bfv\|_2+|\omega|\|\bfv\|_2&\!\!\!\le c\,\lambda_0\,\|\bfu_0\cdot\nabla\bfv+\bfv\cdot\nabla\bfu_0\|_2\\ 
&\!\!\!\le c\,\lambda_0\left(\|\bfu_0\|_{4}\|\nabla\bfv\|_4+\|\nabla\bfu_0\|_{\frac{12}{5}}\|\bfv\|_{12}\right)\,.
\ea$$
Using in the latter the following Nirenberg-type inequalities (see \cite[Theorem 2.1]{CrMa})
$$
\|\nabla\bfv\|_4\le c_0\,\|D^2\bfv\|^{\frac78}_2\|\bfv\|_2^{\frac18}\,,\ \ \|\bfv\|_{12}\le c_0\,\|D^2\bfv\|^{\frac89}_2\|\bfv\|_2^{\frac19}\,,
$$
we infer, with the help of Young's inequality, that
\be
\|D^2\bfv\|_2+|\omega|^{\frac12}\|\nabla\bfv\|_2+|\omega|\|\bfv\|_2\le 
m\,\|\bfv\|_2
\eeq{4.16}
where
$$
m:=c_1\,\left(\lambda_0^8\,\|\nabla\bfu_0\|_{4}^8+\lambda_0^9\|\nabla\bfu_0\|_{\frac{12}{5}}^9\right)
\,,
$$
and $c_1=c_1(\Omega)$.
The desired result follows from \eqref{4.16}  by choosing $\bar\omega:=\max\{m,\lambda_0^2,1\}$.
\QED

We now turn our focus to the study of  some properties of the {\em time-dependent} operator
\be
\mathscr Q:=\omega_0\,(\cdot)_{\tau}+\mathscr L_2:\ \mathscr W_{2\pi,0}^{2}(\Omega)\mapsto \calh_{2\pi,0}(\Omega)\,, 
 \ \ \omega_0>0\,.
\eeq{2.17}
We begin to recall the following result,  proved  in \cite[Lemma 5]{GaA} for the two-dimensional case. However the proof carries over {\em verbatim} to the three-dimensional case and, therefore, will be omitted. 
\Bl The operator
$$
\omega_0\,(\cdot)_{\tau}-{\rm P}\,[\Delta+\lambda_0\,\partial_1]:  \mathscr W_{2\pi,0}^{2}(\Omega)\mapsto \mathscr H_{2\pi,0}(\Omega) 
$$
is a homeomorphism.
\EL{2.4}

With the help of this result, we can prove the following one.
\Bp Let $\bfu_0\in X^{2,\frac43}(\Omega)$. Then, the operator $\mathscr Q$ defined in \eqref{2.17}
is Fredholm of index 0.
\EP{2.5}
{\em Proof.} In view of \lemmref{2.4}, it is enough to show that the operator
$$
\mathscr C:\bfv\in \mathscr W_{2\pi,0}^{2}(\Omega)\mapsto \bfu_0\cdot\nabla\bfv+\bfv\cdot\nabla\bfu_0\in \mathscr L_{2\pi,0}^2(\Omega)
$$
is compact. Let $\{\bfv_k\}\subset \mathscr W_{2\pi,0}^{2}(\Omega)$ with $\|\bfv_k\|_{\mathscr W_{2\pi,0}^{2}}=1$, for all $k\in\nat$. We may then select a sequence (again denoted by $\{\bfv_k\}$) and find $\bfv_*\in \mathscr W_{2\pi,0}^{2}(\Omega)$ such that
\be
\bfv_k\to{\bfv_*} \ \ \mbox{weakly in $\mathscr W_{2\pi,0}^{2}(\Omega)$.}
\eeq{2.18}
Without loss of generality, we may take $\bfv_*\equiv\0$. From \eqref{2.18}, \eqref{4.11}$_2$, and Lions-Aubin lemma we then have
\be
\int_{-\pi}^{\pi}\left(\|\bfv_k(\tau)\|_{12,\Omega_R}^2+\|\nabla\bfv_k(\tau)\|_{4,\Omega_R}^2\right)\to 0\ \ \mbox{as $k\to\infty$, for all $R>R_*$\,.}
\eeq{2.19}
By the H\"older inequality, 
$$
\int_{-\pi}^{\pi} \|\bfu_0\cdot\nabla\bfv_k(\tau)\|_{2}^2
\le \|\bfu_0\|_4\int_{-\pi}^{\pi} \|\nabla\bfv_k(\tau)\|_{4,\Omega_R}^2+ \|\bfu_0\|_{4,\Omega^R}^2\int_{-\pi}^{\pi}\|\nabla\bfv_k(\tau)\|_{4}^2\,,
$$
which, by (4.11)$_1$, \eqref{2.18}, \eqref{2.19}  and the arbitrariness of $R$ furnishes
\be
\lim_{k\to\infty}\int_{-\pi}^{\pi} \|\bfu_0\cdot\nabla\bfv_k(\tau)\|_{2}^2=0\,.
\eeq{2.21}
Likewise, again by H\"older inequality,
$$\ba{rl}\medskip
\Int{-\pi}{\pi} \|\bfv_k(\tau)\cdot\nabla\bfu_0\|_{2}^2
\le &\!\!\!\!\|\nabla\bfu_0\|_{\frac{12}{5}}^2\Int{-\pi}{\pi} \|\bfv_k(\tau)\|_{12,\Omega_R}^2\\ &+ 
\|\nabla\bfu_0\|_{\frac{12}{5},\Omega^R}^2\Int{-\pi}{\pi}\|\bfv_k(\tau)\|_{12}^2\,.
\ea
$$
From the latter, and again (4.11)$_1$, \eqref{2.18}, and \eqref{2.19} we deduce
\be
\lim_{k\to\infty}\int_{-\pi}^{\pi} \|\bfv_k(\tau)\cdot\nabla\bfu_0\|_{2}^2=0\,.
\eeq{2.22}
Combining \eqref{2.21} and \eqref{2.22} we thus conclude
$$
\lim_{k\to\infty}\|\mathscr C(\bfv_k)\|_{L^2(\Omega_{2\pi})}=0\,,
$$
which completes the proof of the lemma.\QED

Our next and final objective is to rewrite \eqref{2.12} in the abstract form \eqref{3.5}, so that under the appropriate assumptions, we may apply \theoref{3.1} and provide the desired bifurcation result.

To that purpose,  we introduce  the scaled time $\tau:=\omega\,t$, split $\bfv$ and ${p}$ as the sum of their time average, $(\bar{\bfv},\bar {p})$, over the time interval $[-\pi,\pi]$, and their ``purely periodic" component $(\bfw:=\bfv-\bar{\bfv},\, \varphi:=\bar{ p}-{p})$. In this way,  problem \eqref{2.12} can be  equivalently rewritten as the following coupled nonlinear elliptic-parabolic problem
\be 
\ba{c}\medskip
\left.\ba{rl}\medskip
\Delta\bar{\bfv}+\lambda_0\big(\,\partial_1\bar\bfv-\bfu_0\cdot\nabla\bar\bfv-\bfu_0\cdot\nabla\bar\bfv\big)&\!\!\!= \nabla \bar{\sf p}+\bfN_1(\bar{\bfv},\bfw,\mu)\\
\Div\bar\bfv&\!\!\!=0\ea\right\}\ \ \mbox{in $\Omega$}\\
\bar\bfv=\0\ \ \mbox{at $\partial\Omega$}\,,\ \ \Lim{|x|\to\infty}\bar\bfv(x)=\0
\ea 
\eeq{3.2_0}
and
\be\ba{cc}\medskip\left.\ba{rl}\smallskip
\omega\,{\bfw}_\tau-\Delta\bfw-\lambda_0\,\big(\partial_1\bfw-&\!\!\!\!\bfu_0\cdot\nabla\bfw-\bfw\cdot\nabla\bfu_0\big)\\ \medskip
&\!\!\!\!=\nabla {\varphi} +\bfN_2(\bar\bfv,\bfw,\mu)\\
\Div\bfw&\!\!\!\!=0
\ea\right\}\ \ \mbox{in $\Omega_{2\pi}$}\\
\quad\quad\bfw=\0\ \ \mbox{at $\partial\Omega_{2\pi}$}\,,\ \ \Lim{|x|\to\infty}\bfw(x,t)=\0\,,\ea
\eeq{3.3_0}
where 
\be\ba{rl}\medskip
\bfN_1:=&\!\!\!\!-\mu\,\left[\partial_1\bar\bfv-\bfu(\mu+\lambda_0)\cdot\nabla\bar\bfv-\bar\bfv\cdot\nabla\bfu(\mu+\lambda_0)\right]\\ \medskip
&\!\!\!\!+\lambda_0\,\big[(\bfu(\mu+\lambda_0)-\bfu_0)\cdot\nabla\bar\bfv+\bar\bfv\cdot\nabla(\bfu(\mu+\lambda_0)-\bfu_0)\big]\\
&\!\!\!\!+(\mu+\lambda_0)\Big[\bar\bfv\cdot\nabla\bar\bfv+
\bar{\bfw\cdot\nabla\bfw}\,\Big]\,,
\ea
\eeq{3.4_0}
and
\be\ba{rl}\medskip
\bfN_2:=&\!\!\!\! \mu\,\left[\partial_1\bfw-\bfu(\mu+\lambda_0)\cdot\nabla\bfw-\bfw\cdot\nabla\bfu(\mu+\lambda_0)\right]\\ \smallskip
&\!\!\!\!-\lambda_0\,\Big[(\bfu(\mu+\lambda_0)-\bfu_0)\cdot\nabla\bfw+\bfw\cdot\nabla(\bfu(\mu+\lambda_0)-\bfu_0)\Big]
\\
&\!\!\!\!+(\mu+\lambda_0)\Big[\bfw\cdot\nabla\bar\bfv+\bar\bfv\cdot\nabla\bfw+\bfw\cdot\nabla\bfw-
\bar{\bfw\cdot\nabla\bfw}\,\Big]\,,
\ea
\eeq{3.5_0}
where, we recall, $\mu:=\lambda-\lambda_0$, and $\bfu_0\equiv \bfu(\lambda_0)$.
\smallskip\par
We prove next some functional properties of the quantities $\bfN_i$, $i=1,2$.
\Bl The following bilinear maps are continuous 
$$\ba{ll}\medskip
\calm_1 :(\bfv_1,\bfv_2)\in [X^{2,\frac43}(\Omega)]^2\mapsto \bfv_1\cdot\nabla\bfv_2\in L^{\frac43}(\Omega)\,,\\ \medskip 
\calm_2: (\bfw_1,\bfw_2)\in[\mathscr W_{2\pi,0}^2(\Omega)]^2\mapsto 
\Int{-\pi}{\pi}
{\bfw_1\cdot\nabla\bfw_2}\in  L^r(\Omega)\,,\ r=\frac43,2\,,\\ \medskip
\calm_3:(\bfv,\bfw)\in X^{2,\frac43}(\Omega)\times \mathscr W_{2\pi,0}^2(\Omega)\mapsto \bfv\cdot\nabla\bfw\in \mathscr L^2_{2\pi,0}(\Omega)\,,\\ \medskip
\calm_4:(\bfv,\bfw)\in X^{2,\frac43}(\Omega)\times \mathscr W_{2\pi,0}^2(\Omega)\mapsto \bfw\cdot\nabla\bfv\in \mathscr L^2_{2\pi,0}(\Omega)\,,\\
\calm_5:(\bfw_1,\bfw_2)\in [\mathscr W_{2\pi,0}^2(\Omega)]^2\mapsto \bfw_1\cdot\nabla\bfw_2\in \mathscr L^2_{2\pi,0}(\Omega)\,.
\ea
$$
\EL{3.1_0}
{\em Proof.} The continuity of $\calm_1$ is shown in \cite[Theorem 2.2]{GaRb}. 
In order to show the remaining properties, we begin to observe that, by  H\"older inequality and \eqref{4.11}, 
$$\ba{ll}\medskip
\|\calm_2(\bfw_1,\bfw_2)\|_{\frac43}\le \Int{-\pi}{\pi}\|\bfw_1\|_{{4}}\|\nabla\bfw_2\|_2\le c_1\,\|\bfw_1\|_{\mathscr W_{2\pi,0}^2}\|\bfw_2\|_{\mathscr W_{2\pi,0}^2}\\ \medskip
\|\calm_2(\bfw_1,\bfw_2)\|_2\le \Int{-\pi}{\pi}\|\bfw_1\|_{4}\|\nabla\bfw_2\|_4\le c_2\,\|\bfw_1\|_{\mathscr W_{2\pi,0}^2}\|\bfw_2\|_{\mathscr W_{2\pi,0}^2}\\  \medskip
\|\calm_3(\bfw,\bfw)\|_{\mathscr L^2_{2\pi,0}}\le (2\pi)^{\frac12}\,\|\bfv\|_4\Big(\Int{-\pi}{\pi}\|\nabla\bfw_2\|_4^2\Big)^{\frac12}\le c_3\,\|\bfv\|_{X^{2,\frac43}}\|\bfw_2\|_{\mathscr W_{2\pi,0}^2}\\
\|\calm_4(\bfw,\bfv)\|_{\mathscr L^2_{2\pi,0}}\le (2\pi)^{\frac12}\|\nabla\bfv\|_{\frac{12}{5}}\Big(\Int{-\pi}{\pi}\|\bfw\|_{12}^2\Big)^{\frac12}\le c_4\,\|\bfv\|_{X^{2,\frac43}}\|\bfw\|_{\mathscr W_{2\pi,0}^2}\,.\ea
$$
Furthermore,
$$\ba{ll}\medskip
\|\calm_5(\bfw_1,\bfw_2)\|_{\mathscr L^2_{2\pi,0}}\le (2\pi)^{\frac12}\,\essup{\tau\in[-\pi,\pi]}\|\bfw_1(\tau)\|_{4}^2\Big(\Int{-\pi}{\pi}\|\nabla\bfw_2\|_{4}^2\Big)^{\frac12}
\\
\hspace*{3.4cm}\le c_5\,\|\bfw_1\|_{\mathscr W_{2\pi,0}^2}\|\bfw_2\|_{\mathscr W_{2\pi,0}^2}\,,\ea$$
where, in the  last step, we have used \eqref{4.11} and the embedding $\mathscr W^2_{2\pi,0}(\Omega)\subset L^\infty(-\pi,\pi;L^4(\Omega))$; see \cite[Theorem 2.1]{Sol}.
\QED

Let
$$\ba{ll}\medskip
\mathscr N_1:(\bar\bfv,\bfw,\mu)\in X_0^{2,\frac43}(\Omega)\times\mathscr W_{2\pi,0}^2(\Omega)\times U(0)\mapsto {\rm P}\,\bfN_1((\bar\bfv,\bfw,\mu)\in H(\Omega)\\ \medskip
\mathscr N_2:(\bar\bfv,\bfw,\mu)\in  X_0^{2,\frac43}(\Omega)\times\mathscr W_{2\pi,0}^2(\Omega)\times U(0)\\
\hspace*{5cm}\mapsto {\rm P}\,\bfN_2(\bar\bfv,\bfw,\mu)\in \mathscr H_{2\pi,0}(\Omega)\,.
\ea
$$
From \lemmref{3.1_0}  it follows that $\mathscr N_i$, $i=1,2$,  are well defined, which allows us to rewrite \eqref{3.2_0}--\eqref{3.5_0} in the following abstract form entirely analogous to \eqref{3.5}, with the obvious interpretation of the function spaces involved:
\be
\mathscr L_1(\bar\bfv)=\mathscr N_1(\bar\bfv,\bfw,\mu)\,\ \mbox{in $H(\Omega)$}\,;\ \ \omega \,\bfw_\tau+\mathscr L_2(\bfw)=\mathscr N_2(\bar\bfv,\bfw,\mu)\,\ \mbox{in $\mathscr H_{2\pi,0}$}\,.  
\eeq{4.26}
Notice that the spatial asymptotic conditions on $\bar\bfv$ and $\bfw$ in \eqref{3.2_0}$_4$ and \eqref{3.3_0}$_4$ are interpreted in the sense of \remref{2.1} and \remref{3.2}. 
Moreover, again by \lemmref{3.1_0} and under the assumptions of \propref{4.1}, we deduce that $\mathscr N_i$, $i=1,2$, are, in fact, analytic in a neighborhood of $(\0,\0,0)\subset X^{2,\frac43}(\Omega)\times\mathscr W_{2\pi,0}^2(\Omega)\times U(0)$. 
We may then show that  $\mathscr N_i$, $i=1,2$, match the assumption (H5) of the abstract formulation, along with the stated properties of the coefficients $R$ and $S$. In particular, it is easy to check that
\be
S_{011}(\bfw)={\rm P}\,\big[\partial_1\bfw-\bfu_0\cdot\nabla\bfw-\bfw\cdot\nabla\bfu_0-\lambda_0\,\big(\bfu^\prime(\lambda_0)\cdot\nabla\bfw+\bfw\cdot\nabla \bfu^\prime(\lambda_0)\big)\big]\,,
\eeq{4.27} 
where ${}^\prime$ means differentiation with respect to $\mu$. 

We now turn to the linear operators $\mathscr L_1$ and $\mathscr L_2$. We assume 
\tag{$\mathcal H1$}
\be
{\sf N}\,[\mathscr L_1]=\{\0\}\,.
\eeq{H1}
Since, by \propref{4.1}, $\mathscr L_1$ is Fredholm of index 0, condition \eqref{H1} implies that (H1) is satisfied.
Furthermore, supported by \propref{4.2}, we assume
\tag{$\mathcal H2$}
\be\ba{ll}\medskip
\nu_0:={\rm i}\,\omega_0 \,\ \mbox{is  an eigenvalue of multiplicity 1 of $\mathscr L_2$}\,,\\ 
k\,\nu_0\,, k\in\nat-\{0,1\}\ \mbox{is not an eigenvalue of $\mathscr L_2$}\,,
\ea
\eeq{H2}
Let $\bfv_1=\Re[\bfv_0\,{\rm e}^{{\rm i}\,\tau}]$, $\bfv_1^*=\Re[\bfv_0\,{\rm e}^{-{\rm i}\,\tau}]$, where $\bfv_0$ and $\bfv_0^*$ are eigenvectors of $\mathscr L_2$ and its adjoint $\mathscr L_2^*$   normalized as in \eqref{3.6} and  corresponding to the eigenvalue $\nu_0$.  Denote by $\nu(\mu)$ the eigenvalue of $\mathscr L_2-\mu\, S_{011}$ with $S_{011}$ given in \eqref{4.27}. By \remref{3.1} we know that $\nu(\mu)$ is a smooth well-defined function and that
$$ 
\Re[{\nu}^\prime(0)]=\big(S_{011}(\bfv_1)|\bfv_1^*\big)\,. 
$$
We then assume
\tag{$\mathcal H3$}
\be
\Re[{\nu}^\prime(0)]\neq 0\,.
\eeq{H3}
Finally, we observe that, thanks to \propref{2.5} the operator $\mathscr Q$ obeys condition (H4). 
\smallskip\par
The following  bifurcation result for the steady-state flow of a Navier-Stokes liquid past an obstacle is then an immediate consequence of \theoref{3.1}.
\Bt Suppose  \eqref{H1}--\eqref{H3} hold. 
Then, the following properties are valid. \smallskip\\
{\rm (a)} {\rm Existence.} There are analytic families\renewcommand{\theequation}{\arabic{section}.\arabic{equation}}
\setcounter{equation}{27}
\be
\big(\bar\bfv(\varepsilon),\bfw(\varepsilon),\omega(\varepsilon),\mu(\varepsilon)\big)\in X^{2,\frac43}_0(\Omega)\times \mathscr W_{2\pi,0}^2(\Omega)\times \real_+\times\real
\eeq{fam_0}
satisfying \eqref{3.2_0}--\eqref{3.5_0}, for all $\varepsilon$ in a neighborhood $\mathcal I(0)$ and such that
$$
\big(\bar\bfv(\varepsilon),\bfw(\varepsilon)-\varepsilon\,\bfv_1,\omega(\varepsilon),\mu(\varepsilon)\big)\to (0,0,\omega_0,0)\ \ \mbox{as $\varepsilon\to 0$}\,.
$$
\par\noindent
{\rm (a)} {\rm Uniqueness.}
There is a neighborhood  $$U(0,0,\omega_0,0)\subset X^{2,\frac43}_0(\Omega)\times \mathscr W_{2\pi,0}^2(\Omega)\times \real_+\times\real$$ such that every (nontrivial) $2\pi$-periodic solution to \eqref{3.2_0}--\eqref{3.5_0}, $(\bfz,\bfs)$, lying in $U$ must coincide, up to a phase shift, with that member of the family \eqref{fam_0} having $\varepsilon\equiv(\bfs|\bfv_1^*)$.
\smallskip\par\noindent
{\rm (a)} {\rm Parity.}  The functions $\omega(\varepsilon)$ and $\mu(\varepsilon)$ are even:
$$
\omega(\varepsilon)=\omega(-\varepsilon)\,,\ \ \mu(\varepsilon)=\mu(-\varepsilon)\,,\ \ \mbox{for all $\varepsilon\in\cali(0)$\,.} 
$$
Consequently, the bifurcation due to these solutions is either subcritical or supercritical, a two-sided bifurcation being excluded.\footnote{Unless $\mu\equiv 0$.}
\ET{4.1}

\ed
\begin{thebibliography}{99}
\bibitem{Bab3} Babenko, K.I., On properties of steady viscous incompressible fluid flows.  Approximation methods for Navier-Stokes problems ({\em Proc. Sympos., Univ. Paderborn, Paderborn, 1979}),  pp. 12--42, Lecture Notes in Math., Vol. 771, Springer, Berlin, 1980
\bibitem{Bab}Babenko, K. I., On the spectrum of a linearized problem of flow of a viscous incompressible fluid around a body (Russian), {\em Dokl. Akad. Nauk SSSR},  {\bf 262} 64--68  (1982)
\bibitem{Bab2}Babenko, K.I., Periodic solutions of a problem of the flow of a viscous fluid around a body, {\em Soviet Math. Dokl.} {\bf 25}  211--216 (1982)
\bibitem{CI} Chossat, P., and  Iooss, G., {\em The Couette-Taylor problem}, Applied Mathematical Sciences, Vol. 102, Springer-Verlag, New York (1994)
\bibitem{CrMa}Crispo, F. and Maremonti, P., An interpolation inequality in exterior
domains, {\em Rend. Sem. Mat. Univ. Padova}, {\bf 112} 11--39 (2004)
\bibitem{GaB}Galdi, G.P., {\em An introduction to the mathematical theory of the Navier-Stokes equations.
Steady-state problems}, Second edition. Springer Monographs in Mathematics,
Springer, New York (2011)
\bibitem{GaA}Galdi, G.P., On time-periodic flow of a viscous liquid past a moving cylinder, {\em Arch. Ration. Mech. Anal.},  {\bf 210} 451--498 (2013)
\bibitem{GaBi}Galdi, G.P., On bifurcating time-periodic flow of a Navier-Stokes liquid past a cylinder,  arXiv:1506.02945
\bibitem{GaRb}Galdi, G.P., and Rabier, P.J., Sharp existence results
for the stationary Navier-Stokes problem
in three-dimensional exterior domains, {\em Arch. Rational Mech. Anal.}, {\bf 154} 343--368 (2000)

\bibitem{GoGoKa}Gohberg, I., Goldberg, S. and Kaashoek, M.A., {\em Classes of linear operators:I. Operator Theory},  Advances and Applications, Vol.49, Birkh\"auser Verlag, Basel (1990)
\bibitem{Gui}Guyon, E., Hulin, J.-P., Petit, L., and Mitescu, C.D., {\em
Physical Hydrodynamics}
Oxford University Press (second Edition) (2015) 
\bibitem{Hopf}Hopf, E., Abzweigung einer periodischen L\"osung von einer station\"aren eines Differentialsystems, {\em Ber. Verh. Sächs. Akad. Wiss. Leipzig. Math.-Nat. Kl.}   {\bf 95},   3--22 (1943)
\bibitem{Io}Iooss, G., Existence et stabilit\'e de la solution p\'eriodiques secondaire intervenant dans les
probl\'emes d'evolution du type Navier-Stokes, {\em Arch. Rational Mech. Analysis}, {\bf 47}, 301--329
(1972)
\bibitem{Yu} Iudovich, V.I., The onset of auto-oscillations in a fluid, {\em J. Appl. Math. Mech.}  {\bf 35}, , 587--603  (1971)
\bibitem{JS}Joseph, D.D., and Sattinger, D.H.,
Bifurcating time periodic solutions and their stability. 
{\em Arch. Rational Mech. Anal.},  {\bf 45}79--109   (1972) 
\bibitem{Kato}Kato, T., {\em Perturbation theory for linear operators}, Springer Classics in Mathematics (1995)
\bibitem{Saz}Sazonov, L.I., The onset of auto-oscillations in a flow,  {\em Siberian Math. J.}  {\bf 35}    1202--1209 (1994)
\bibitem{Sol}Solonnikov, V.A.,
     \newblock {Estimates of the solutions of the nonstationary Navier-Stokes system},
     \newblock {\em Zap. Naucn. Sem. Leningrad. Otdel. Mat. Inst. Steklov} (LOMI), \textbf{38} 153--201 (1973)
\bibitem{Tritton}Tritton, D.J., {\em Physical Fluid Dynamics}, Second Edition, Clarendon Press, Oxford (1988)
\bibitem{Z} Zeidler, E., {\em Nonlinear Functional Analysis and Applications}, Vol.1, Fixed-Point Theorems, Springer-Verlag, New York (1986) 
\bibitem{Z1}Zeidler, E., {\em Nonlinear Functional Analysis and Applications}, Vol.4, Application to Mathematical Physics, Springer-Verlag, New York  (1988) 

\end{thebibliography}
